\documentclass[10pt, sigconf, table, ]{acmart}

\usepackage{graphicx,balance,multirow,xspace,subcaption,longtable}
\usepackage{makecell,amsmath,amssymb,cleveref,rotating}
\usepackage[font={small}, textfont=md]{caption}
\usepackage[T1]{fontenc}
\usepackage[draft,inline,nomargin,index]{fixme}
\usepackage{booktabs}

\PassOptionsToPackage{hyphens}{url}
\usepackage{hyperref}
\newcolumntype{P}[1]{>{\arraybackslash}p{#1}}
\newcolumntype{X}[1]{>{\centering\arraybackslash}p{#1}}

\expandafter\def\expandafter\UrlBreaks\expandafter{\UrlBreaks
  \do\a\do\b\do\c\do\d\do\e\do\f\do\g\do\h\do\i\do\j%
  \do\k\do\l\do\m\do\n\do\o\do\p\do\q\do\r\do\s\do\t%
  \do\u\do\v\do\w\do\x\do\y\do\z\do\A\do\B\do\C\do\D%
  \do\E\do\F\do\G\do\H\do\I\do\J\do\K\do\L\do\M\do\N%
  \do\O\do\P\do\Q\do\R\do\S\do\T\do\U\do\V\do\W\do\X%
  \do\Y\do\Z}

\crefformat{section}{\S#2#1#3}
\crefformat{subsection}{\S#2#1#3}
\crefformat{subsubsection}{\S#2#1#3}

\fxsetup{theme=colorsig,mode=multiuser,inlineface=\itshape,envface=\itshape}
\FXRegisterAuthor{hh}{ahh}{Hussam}
\FXRegisterAuthor{mm}{amm}{Maaz}
\FXRegisterAuthor{fz}{afz}{Fareed}
\FXRegisterAuthor{rn}{arn}{Rishab}

\hypersetup{%
  pdftitle={Act or React?},
  pdfauthor={},
  pdfkeywords={},
  bookmarksnumbered,
  pdfstartview={FitH},
  colorlinks,
  citecolor=black,
  filecolor=black,
  linkcolor=black,
  urlcolor=linkColor,
  breaklinks=true,
  hypertexnames=false
}

\newcommand\clearrow{\global\let\rowmac\relax}

\newcommand{\para}[1]{{\vspace{.05in} \bf \noindent #1 }}

\newcommand{\bannedsubreddit}[2]{\emph{r/#1 (banned in #2)}}

\newcommand{\activesubreddit}[1]{\emph{r/#1 (active as of 4/2019)}}
\newcommand{\subreddit}[1]{\emph{r/#1}}

\newcommand{\red}[1]{\textcolor{red}{\bf #1}}
\newcommand{\blue}[1]{\textcolor{blue}{\bf #1}}

\renewcommand{\footnotesize}{\scriptsize}
\newcommand{\etc}{etc.}
\newcommand{\eg}{e.g.,\ }
\newcommand{\etal}{et al.\xspace}
\newcommand{\ie}{i.e.,\ }

\setcopyright{none}

\begin{document}

\renewcommand\footnotetextcopyrightpermission[1]{} 
\pagestyle{plain} 

\renewcommand{\sectionautorefname}{\S}
\renewcommand{\subsectionautorefname}{\S}
\renewcommand{\subsubsectionautorefname}{\S}

\title{To Act or React?}
\subtitle{Investigating Proactive Strategies For Online Community Moderation}

\author{Hussam Habib}
\author{Maaz Bin Musa}
\author{Fareed Zaffar}
\author{Rishab Nithyanand}

\renewcommand{\shortauthors}{H. Habib \etal}

\begin{abstract}

  Reddit, the self-proclaimed ``front page of the Internet'' with over 330M
  active users, has found its communities playing a prominent role in
  originating and propagating sexist, racist, and hateful socio-political
  discourse. Reddit administrators have generally struggled to prevent or
  contain such discourse for several reasons including: (1) the inability for
  a handful of human administrators to track and react to millions of posts and
  comments per day and (2) fear of backlash as a consequence of administrative
  decisions to ban or quarantine hateful communities. Consequently, as shown in
  our background research, administrative actions (community bans and
  quarantines) are often taken in reaction to media pressure following
  offensive discourse within a community spilling into the real world with
  serious consequences.  
  In this paper, we investigate the feasibility of proactive moderation on
  Reddit -- \ie proactively identifying communities at risk of committing
  offenses that previously resulted in bans for other communities. Proactive
  moderation strategies show promise for two reasons: (1) they have potential
  to narrow down the communities that administrators need to monitor for
  hateful content and (2) they give administrators a scientific rationale to
  back their administrative decisions and interventions. Our work shows that
  communities are constantly evolving in their user base and topics of
  discourse and that evolution into hateful or dangerous (\ie considered
  bannable by Reddit administrators) communities can often be predicted months
  ahead of time. This makes proactive moderation feasible. Further, we leverage
  explainable machine learning to help identify the strongest predictors of
  evolution into dangerous communities. This provides administrators with
  insights into the characteristics of communities at risk becoming dangerous
  or hateful. Finally, we investigate, at scale, the impact of participation in
  hateful and dangerous subreddits and the effectiveness of community bans and
  quarantines on the behavior of members of these communities. 
  
\end{abstract}

\maketitle

\sloppy

\section{Introduction}\label{sec:introduction}

Reddit, the self-proclaimed \emph{``front page of the internet''}, has over
138K active communities, called \emph{subreddits}, with over 330M  active
users \cite{Reddit-Web2019}. 
In recent years, the site has been mired in controversies around the
role that its communities played in originating and propagating sexist,
racist, and generally hateful online socio-political discourse. A few of the
recent controversies have involved communities such as:
\bannedsubreddit{Physical\_Removal}{8/2017} which advocated for the physical
removal of liberals in the United States prior to and even after the murder of
Heather Heyer in Charlottesville \cite{HeatherHeyer-DailyBeast2017},
\bannedsubreddit{incels}{11/2017} which endorsed and celebrated the murder of
and violence against sexually active women \cite{AlekMinassian-NBCNews2018},
\bannedsubreddit{greatawakening}{3/2018} and
\bannedsubreddit{pizzagate}{11/2016} which falsely alleged the existence of
child trafficking rings by the US Democratic Party and left-wing corporations
resulting in real-life attacks, threats, and harrassment (\eg against the Comet
Ping Pong restaurants \cite{Pizzagate-WaPo2016}),
\bannedsubreddit{WatchPeopleDie}{3/2019}, \bannedsubreddit{gore}{3/2019}
which disseminated videos of the Christchurch Mosque shootings
\cite{Christchurch-TheVerge2019}, and \activesubreddit{The\_Donald} which
continues to pedal white genocide conspiracy theories
\cite{TheDonald-WhiteGenocide-SPLC2018}, false-flag theories in the wake of
shootings and bomb threats \cite{TheDonald-FalseFlags-NYT2018}, and violent
anti-immigrant \cite{TheDonald-AntiImmigrant-FC2018} and anti-Islam rhetoric
\cite{TheDonald-AntiIslam-MotherJones2019}. 

In reaction to many of these controversies, Reddit has resorted to banning or
quarantining subreddits citing violations of the Reddit content policy
\cite{Reddit-ContentPolicy2019} which prohibits specific types of content
including content which ``\emph{encourages or incites violence}''.
However, the effectiveness and timeliness of such bans and quarantines is
frequently debated. While previous research \cite{Chandrasekharan-CSCW2017}
concluded that such bans ``\emph{worked for Reddit}'', others have pointed out
that they are too reactionary and occur only after a significant amount of
damage has already been observed \cite{RedditModeration-Mashable2019,
RedditModeration-Vox2017, Marantz-NY2018}. Along another dimension, Reddit has
also faced criticism for inconsistent and seemingly ad-hoc applications of the
content policy by those claiming that the platform provides a safe-haven for
extremist ideologies and others claiming that the platform leverages the
content policy as a mechanism to censor ``non-mainstream'' opinions and
ideologies.

Despite these arguments about how Reddit should (not) be moderated, little is
actually known about how subreddits evolve over time, the evolutionary
predictors of offensive or dangerous subreddits, and how moderation decisions
impact the wider community. This gap in knowledge presents Reddit
administrators with several challenging questions: {how to identify offensive
or dangerous subreddits} before serious harm has been caused by them, how to
moderate the impact of these subreddits on users and other subreddits, and how
to scientifically rationalize these moderation decisions. In this paper, we
seek to answer these questions. We do so by setting out to test the following
hypotheses.

  \para{H1. Subreddits may not converge to topical or user base stability.
    (\Cref{sec:evolution})}
    This hypothesis, if valid, will show that subreddits need to be constantly
    monitored for changing discourse and moderation decisions need to be
    re-evaluated from time to time -- an expensive proposition for a small
    number of human moderators and administrators. If invalid, the failed
    hypothesis will show that subreddits always converge to topical and
    user base stability and therefore only need to be evaluated by moderators
    once -- after topical or community stability has been reached.
    In order to test this hypothesis, we develop techniques to track the nature
    and magnitude of subreddit evolution. These techniques permit us to
    quantify the distance between two subreddits and also identify subreddits
    with similar evolutionary behaviors.  

  \para{H2. Evolution into hateful or dangerous subreddits can be predicted.
    (\Cref{sec:predictors})}
    This hypothesis, if valid, will show that tools may be built to help
    moderators pre-emptively identify subreddits likely to devolve into hateful
    or dangerous subreddits. If invalid, the failed hypothesis will show that
    moderators cannot perform pre-emptive actions to mitigate the impact of
    offensive or dangerous subreddits.
    In order to test this hypothesis, we develop explainable machine learning
    techniques to identify the value of community-, user-, moderator- and
    structure-based features in predicting the evolutionary outcome of
    a subreddit. 
   
  \para{H3. User participation in a hateful subreddit negatively
    changes the nature of their participation within the broader community and
    this can be corrected by bans or quarantines. (\Cref{sec:moderation})}
    This hypothesis, if valid, will show that banning or quarantining hateful
    subreddits is an effective way to curb their impact on the broader
    community. If invalid, the failed hypothesis will point to the need for
    more nuanced techniques to moderate the impact of hateful communities.
    In order to test this hypothesis, we perform large-scale analysis of the
    impact of join, community ban, and quarantine events on user behavior.

\section{Reddit: The Platform and Dataset}\label{sec:background}

Reddit is currently the sixth most popular site in the USA with over 330M
active monthly users \cite{Reddit-Web2019}. In recent years, Reddit
has come under increasing criticism for the types of content shared by its
users \cite{TheDonald-WhiteGenocide-SPLC2018,
TheDonald-FalseFlags-NYT2018, TheDonald-AntiImmigrant-FC2018,
TheDonald-AntiIslam-MotherJones2019}. This has resulted in numerous changes of
the Reddit content policy and moderation strategies. In this section, we
provide a high-level overview of Reddit with a focus on its content moderation
policies (\Cref{sec:background:overview}) and our datasets
(\Cref{sec:background:data}).

\begin{table}[t]
 \centering
 \small
 \begin{tabular}{p{3.15in}}
  \toprule
  \textbf{Content is prohibited if it}\\\midrule
  is illegal\\
  is involuntary pornography \\
  is sexual or suggestive content involving minors \\
  encourages or incites violence \\
  threatens, harasses, or bullies or encourages others to do so \\
  is personal and confidential information \\
  impersonates someone in a misleading or deceptive manner \\
  uses Reddit to solicit or facilitate any transaction gift involving certain goods and services \\
  is spam \\
  \bottomrule
 \end{tabular}
 \caption{Prohibited content according to Reddit's content policy.
 \cite{Reddit-ContentPolicy2019}}
 \label{tab:background:contentpolicy}
\end{table}

\subsection{An overview of the Reddit platform}\label{sec:background:overview}

\begin{table*}[t]
  \centering
  \small
  \begin{tabular}{c p{2.75in} p{2.75in}}
    \toprule
    \textbf{Month/Year} & \textbf{Event} & \textbf{Admin actions}\\
    \midrule
   %
   %
   %
    02/2015
    & \subreddit{TheFappening} gains media attention for facilitating
    distribution of leaked celebrity nudes. \cite{Fappening-Media2014-1,
    Fappening-Media2014-2, Fappening-Media2014-3}
    & Content policy amended to prohibit ``involuntary pornography'' and
    \subreddit{TheFappening} is banned citing this policy.
    \cite{Fappening-Reddit2014-1, Fappening-Reddit2015-1}\\

    11/2016
    & \subreddit{pizzagate} gains media attention after the Comet Ping Pong
    restaurant in Washington DC begins receiving threats in response to
    the conspiracy alleging the existence of a child trafficking operation by
    Democratic politicians. \cite{Pizzagate-WaPo2016, Pizzagate-Media2016-1} 
    & \subreddit{pizzagate} is banned citing policy about ``posting personal
    and confidential information'' and ``threatening, harassing, and bullying
    others''. \cite{Pizzagate-WaPo2016}\\

    08/2017
    & \subreddit{Physical\_Removal} gains media attention for advocating and 
    celebrating violence against ``liberals'' and Democrats in the wake of
    the death of Heather Heyer in Charlottesville. \cite{PR-Media2017-1}.
    & \subreddit{Physical\_Removal} is banned citing policy about
    ``threatening, harassing, and bullying others''. \cite{PR-Reddit2017-1,
    PR-Reddit2017-2} \\

    10/2017 
    & Prolific \subreddit{The\_Donald} user is charged with murder of father
    over dispute about participation in Nazi communities.
    \cite{Nazi-Media2017-1}
    & Content policy amended to prohibit ``encouraging or inciting violence''
    and many subreddits including \subreddit{Nazi},
    \subreddit{EuropeanNationalism}, and \subreddit{NationalSocialism} are
    banned citing this new policy. \cite{Nazi-Reddit2017-1, Nazi-Reddit2017-2}
    \\

    11/2017
    & \subreddit{incels} gains media attention due to the subreddit's
    encouragement of violence against women. \cite{Incels-Media2017-1,
    Incels-Media2017-2, Incels-Media2017-3}
    & \subreddit{incels} is banned citing policy about ``encouraging or
    inciting violence''. \cite{Incels-Reddit2017-1, Incels-Reddit2017-2}\\

    02/2018
    & \subreddit{Deepfakes} gains media attention for facilitating the
    distribution of AI generated pornography involving popular actresses.
    \cite{DF-Media2018-1, DF-Media2018-2, DF-Media2018-3}
    & \subreddit{Deepfakes} is banned. Content policy amended to include
    deepfakes in ``involuntary pornography''. \cite{DF-Reddit2018-1,
    DF-Reddit2018-2}\\

    03/2018
    & Parkland Florida school shooting brings debates on gun violence to the
    forefront and articles from 2014 which highlight the thriving gun trade on
    Reddit begin to emerge. \cite{Guns-Media2014-1} 
    & Content policy amended to prohibit ``transactions involving certain
    goods'' and many subreddits including \subreddit{GunDeals} and
    \subreddit{GunsForSale} are banned citing this new policy.
    \cite{Guns-Reddit2018-1}\\

    04/2018
    & \subreddit{Braincels}, a spin-off of the banned \subreddit{incels}
    community, gains media attention for praising the actions of Alek Minassian
    -- alleged perpetrator of the Toronto van attacks.
    \cite{Braincels-Media2018-1, Braincels-Media2018-2, Braincels-Media2018-3}
    & \subreddit{Braincels} and \subreddit{Mindcels} are banned while
    \subreddit{TheRedPill} is quarantined for ``encouraging and inciting
    violence''. \cite{Braincels-Reddit2018-1} \\

    03/2019
    & \subreddit{gore} and \subreddit{WatchPeopleDie} gain media attention for
    facilitating the distribution of videos of the Christchurch shootings.
    \cite{NZ-Media2019-1, NZ-Media2019-2}
    & \subreddit{gore} and \subreddit{WatchPeopleDie} are banned citing policy
    about ``encouraging or inciting violence''. \cite{NZ-Reddit2019-1}\\

    \bottomrule
  \end{tabular}
  \caption{Reactionary administration: A sample of recent events which resulted
  in administrative changes to content policy and reactionary bans of
  subreddits long known to violate the content policy.}
  \label{tab:background:reactionary}
\end{table*}

\para{Redditors and subreddits.} At a high-level, Reddit is a content
aggregation platform where users, also called \emph{redditors}, share content
on topical forums called \emph{subreddits}. Subreddits are generally formed
around specific topics and can range from broad (\eg \subreddit{politics}
which focuses on US politics) to extremely niche (\eg \subreddit{birdswitharms}
which focuses on photoshopped images of birds with human arms). There are
currently 138K active subreddits \cite{Reddit-Web2019} which contain content
posted by redditors. In addition to posting content, redditors can also
interact with each other by commenting on posts and replying to other comments.

\para{Democratized content-curation.} Unlike other social and aggregation
platforms, Reddit relies on its users for more than content generation and
propagation. Redditors also play the role of content curators. Redditors may
also curate content by up- or down-voting comments and posts. Content (comments
or posts) with a high net-vote total is, by default, given very high
visibility. For example, the comments (on a post) and posts (in a subreddit)
are, by default, ordered by decreasing net-vote total. This mechanism lets
users decide which comments and posts are most visible to the rest of the
community and which contributions are silenced or hidden. 

\para{Decentralized moderation.} Besides letting every redditor curate
content by way of voting, Reddit also allows its users to create and moderate
subreddits. Subreddit moderators typically choose their own fellow moderators
from within the community, with a few exceptions for newly created communities
and cases where there are no volunteers within the community. 
Subreddit moderators are tasked with setting and enforcing the rules of
engagement within a subreddit. Moderators may enforce rules via the use of user
bans and content deletion. However, the actions performed by the subreddit
moderators do not impact redditors outside of that subreddit (\eg a subreddit
moderator cannot enforce site-wide bans). 

\para{Centralized administration.} In addition to relying on volunteering
community members to moderate their own subreddits, Reddit also employs
administrators to set and enforce site-wide policies for content and user
engagement. These policies are mandatory and applied \emph{in addition to}
a subreddit's own policies. Administrators have the ability to: (1)  ban users
from making posts or comments visible to the rest of the platform (this is
referred to as a \emph{shadow ban}), (2) prevent subreddits from appearing on
the Reddit front-page and in search results (this is referred to as a
\emph{quarantine}), and (3) ban subreddits from the platform. While these
actions are available for use in response to egregious or repeated violations
of the Reddit content policy, they are rarely applied except in reaction to
media pressure or catastrophic external events occurring as a result of
discourse on the platform. Instead, Reddit administrators encourage users
unhappy with certain communities to create their own communities which enforce
their preferred rules and norms \cite{Auerbach-hunt}. This policy of
reactionary administration has been the subject of much criticism in the wake
of recent controversies on Reddit.

\para{The reactionary content policy.} The Reddit content policy defines
content which is acceptable for posting on the platform. Failing to adhere to
this policy may result in the moderator or administrator actions described
above. Prohibited content according to the April 2019 content policy is shown
in \Cref{tab:background:contentpolicy}. 
In addition to prohibiting content,
Reddit also regulates content containing nudity, pornography, or profanity by
requiring them to be tagged as \emph{Not Safe For Work (NSFW)}.  
Additions to the policy have been made, in large part, in reaction
to negative media coverage of certain events on the platform. 
We illustrate
several of these events and the reactionary administrative actions caused by
them in \Cref{tab:background:reactionary}.

\begin{table}[t]
  \centering
  \small
  \begin{tabular}{l l l l l p{1.25in}}
    \toprule
    \textbf{Dataset} & \textbf{$N_{s}$} & \textbf{$N_{u}$} & \textbf{$N_{p}$} & \textbf{$N_{c}$} & \textbf{Category} \\\midrule
    $\mathfrak{D}_A$ & 3K  & 22M & 203M & 3B & Most active. \\
    $\mathfrak{D}_B$ & 38  & 1M & 5M & 27M & Banned or quarantined. \\
    $\mathfrak{D}_H$ & 118 & 4M & 9M & 141M & Hateful. \\
    $\mathfrak{D}_R$ & 152 & 7M & 16M & 353M & Related to $\mathfrak{D}_H$ and $\mathfrak{D}_B$.\\
    \bottomrule
  \end{tabular}
  \caption{Summary of datasets analyzed in this paper. $N_s$ is the number of
  subreddits in the dataset, $N_u$ is the number of unique users observed,
  $N_p$ is the number of posts, and $N_c$ is the number of comments in the
  selected subreddits. All datasets only contain activity between 01/2015 and
  10/2018.}
  \label{tab:background:data}
\end{table}

\subsection{The Reddit dataset}\label{sec:background:data}

In total, Reddit consists of over 3.5B comments on over 300M posts from over
30M unique users on 1.2M subreddits over the period from 2007-2019. In this
paper, however, we focus on a subset of the entire platform. Specifically, we
perform all the analysis shown in the remainder of this paper on four specific
categories of subreddits described below. These are also summarized in
\Cref{tab:background:data}. All the datasets studied in this work were gathered
using the publicly available Reddit BigQuery dataset
\cite{RedditBQ2, RedditBQ1}.

\para{Most active subreddits ($\mathfrak{D}_A$).} We select the 3K subreddits
which on average had the most number of user posts per month, during the period
from 01/2015 - 10/2018. These subreddits account for 6\% of all posts and 11\%
of all comments made on Reddit during this period. We rely on this dataset to
understand the evolution and life-cycle of a \emph{typical} popular subreddit.

\para{Banned or quarantined subreddits ($\mathfrak{D}_B$).} We select 38
subreddits which were banned or quarantined during the period from 01/2015
- 10/2018. Examples of subreddits in this group include:
\bannedsubreddit{Physical\_Removal}{8/2017} and
\bannedsubreddit{WatchPeopleDie}{3/2019}. We use this dataset to understand
the evolution and life-cycle of subreddits confirmed to violate the
content policy.

\para{Hateful subreddits ($\mathfrak{D}_H$).} We select 118 subreddits that have
been frequently reported by redditors and the media for violating the Reddit
content policy, yet have not been banned or quarantined by administrators. This
list is compiled by analyzing the \subreddit{againsthatesubreddits} and
\subreddit{SubredditDrama} to identify the most frequently user-reported
subreddits. Additional subreddits are manually added to this list based on
media reports. Examples of subreddits in this category include
\subreddit{metacanada} and \subreddit{KotakuInAction}. This dataset was used to
understand the evolution and life-cycle of frequently reported subreddits.

\para{Subreddits related to $\mathfrak{D}_H$ and $\mathfrak{D}_B$
($\mathfrak{D}_R$).} Finally, we extracted subreddits deemed to be
\emph{similar} to those in $\mathfrak{D}_H$ and $\mathfrak{D}_B$.
Similarity between subreddits was measured by computing the cosine similarity of the
latent vectors associated with each subreddit. We explain this process in more
detail in \Cref{sec:evolution:methods}. In total, this dataset contains 152
subreddits including \subreddit{AskTrumpSupporters} and \subreddit{SocialJusticeInAction}. This
dataset was used to understand the evolution and life-cycle of subreddits
likely to share similar characteristics to banned and dangerous subreddits.

\para{Data preprocessing.} In this work, we seek to understand how subreddit
characteristics evolve and if these characteristics can be predictors of
dangerous behavior or hatefulness. Therefore, we need to consider how
subreddit states change at different points in time. We achieve this by
breaking down our data for each subreddit into one-month slots (starting from
1/2015 until 10/2018) and individually analyzing the characteristics of these
``\emph{subreddit states}''.

\section{Subreddit Evolution and Convergence}\label{sec:evolution}

In this section, we focus on testing the following hypothesis:  \textbf{H1.
Subreddits may not converge to topical or user-base stability}. If valid, this
hypothesis demonstrates: (1) the need for techniques which can monitor
subreddit evolution and frequently evaluate the suitability of making
moderation decisions and (2) the potential for identifying evolutionary
patterns that can serve as early predictors for subreddits likely to require
future administrator interventions for offensiveness. If invalid, the failed
hypothesis will show that subreddits converge to topical or user base stability
and therefore only need to be evaluated by moderators once -- after stability
has been reached.
Our methods and results are outlined in \Cref{tab:evolution}.

\begin{table}
  \centering
  \small
  \begin{tabular}{p{2.7in}l}
    \toprule
    \textbf{Methodological questions} & \textbf{\Cref{sec:evolution:methods}}\\
    \midrule
    How do we convert \emph{subreddit states} into fixed-length vectors
    representing topics and active users? & \Cref{sec:evolution:methods:lsa} 
    \\
    How do we compute the topical or user base \emph{distance} between two
    subreddit states? & \Cref{sec:evolution:methods:clustering}
    \\
    \midrule 
    \textbf{Research questions} & \textbf{\Cref{sec:evolution:results}}\\
    \midrule
    How much do subreddit topics and user bases evolve per month on average?
    & \Cref{sec:evolution:results:distributions} \\
    How do subreddit topics and user bases evolve as a function of subreddit age?
    & \Cref{sec:evolution:results:age} \\
    Do subreddits converge to topical or user base stability? 
    & \Cref{sec:evolution:results:takeaway} \\
    \bottomrule
  \end{tabular}
  \caption{Hypothesis 1: Subreddits may not converge to topical or user
  stability. Summary of methods and results.}
  \label{tab:evolution}
\end{table}

\para{High-level overview.} Our goal is to identify if subreddits reach
topical or user stability -- \ie if the topics of discussion and users
participating in these discussions stabilize over time. In order to accomplish
this goal, we need a technique to quantify the (dis)similarity in topics and
user bases for a subreddit at two points in time. 
We accomplish this using the following approach, which is explained in more
detail in \Cref{sec:evolution:methods}. First, for each month of activity in
each subreddit (\ie for each \emph{subreddit state}), we generate a pair of
vectors --  a vector representative of the topics being discussed in posts and
comments in the subreddit (\ie a topic vector) and a vector representative of
active users in the subreddit (\ie an active user vector)
using Latent Similarity Analysis (LSA) \cite{landauer1998introduction}. Second,
we perform hierarchical clustering on both types of vectors across all the
subreddit states in our dataset. We then quantify the \emph{evolutionary
distance} of topics or active users between any two states of a specific
subreddit as the \emph{height of the nearest common parent} in our topic or
active user clustering model, respectively. 

We claim that the topics or user base of a subreddit ($s$) has \emph{converged}
between times $t_1$ and $t_2$ if the evolutionary distance between the
corresponding vectors at these timestamps is minimum. We present our results,
which broadly (\ie for all categories of subreddits in our study) validate the
proposed hypothesis in \Cref{sec:evolution:results}.

\subsection{Methods}\label{sec:evolution:methods}

Our method for testing the hypothesis involves two steps: First, we
create summaries, in the form of latent vectors, of topics and user
participation for each subreddit state (\Cref{sec:evolution:methods:lsa}).
Next, we measure the similarities of these summaries as a function of time
(\Cref{sec:evolution:methods:clustering}).

\subsubsection{How do we convert subreddit states into fixed-length vectors
representing topics and active users?}\label{sec:evolution:methods:lsa}

We generate two types of latent vectors for each subreddit state -- a topic
vector and an active user vector. These are generated as follows.

  \para{Topic vectors.} For each subreddit state in our dataset, we
    randomly sampled 10\% of all comments and used these to generate topic
    vectors. A 10\% sample was used to mitigate the infeasibility of
    efficiently processing billions of comments. Each sampled comment was
    pre-processed by removing English-language stop words, tokenizing, and
    stemming. A count vector, which keeps count of the number of occurrences of
    each tokenized and stemmed word, was then generated for each subreddit
    state. With these count vectors, we created a matrix $M$ in which each row
    is associated with one subreddit state in our dataset and each column is
    associated with a unique stemmed token. The cell $M_{ij}$ represents the
    number of times the $j^{th}$ token was observed in our sampled comments
    obtained from the $i^{th}$ subreddit state. However, since simple frequency
    counts are not discriminative, we follow standard recommendations from
    Jurafsky and Martin \cite{jurafsky2000speech} and apply PPMI (Positive
    Pointwise Mutual Information) weighting to each word. 
    PPMI \cite{bullinaria2007extracting, fano1963transmission} draws on the
    intuition that the association between two words should be weighted by the
    difference of how often they actually co-occur in comparison to how much we
    would expect them to co-occur if they were truly independent. We then view
    each row as the topic vector associated with a subreddit state.


  \para{Active user vectors.} We used a similar approach as above to
    encode the active participants in each subreddit state with only a few
    minor changes. First, for each subreddit state in our dataset, we
    identified the set of unique and active user observed. Second, we selected
    a subset of \emph{reference} subreddit states which served as the reference
    points used to compare all other subreddits -- \ie we considered how all
    other subreddit states appeared with respect to these reference points. In
    our study, we selected all subreddits with between 5K-15K active users as
    the reference set. Next, we created an active user co-occurrence matrix
    where $M_{ij}$ represented the number of common active users between
    subreddit state $i$ and reference subreddit state $j$. Similar to our
    creation of topic vectors, we applied PPMI on this raw co-occurrence matrix
    to uncover information about highly co-occurring cohorts. Finally, since
    our matrix was sparse and high-dimensional, we performed dimensionality
    reduction by running PCA (Principle Component Analysis) and selected
    the top 750 principle components. We selected 750 components since it
    offered the best trade-off between dimension reduction and dataset variance
    loss -- \ie we only saw a $<$ 1\% reduction in dataset variance and 62.5\%
    reduction in dimensions. We then view selected row in this 750-dimension
    matrix as the active user vector associated with a subreddit state.

We note that both the methods described above essentially create
representations of different subreddit states in the same $n$-dimensional
space. 
A similar approach was used in previous work seeking to uncover the
commonalities between \subreddit{The\_Donald} and other communities
\cite{538-DissectingDonald} using vector algebra.

\subsubsection{How do we compute the evolutionary distance between two
subreddit states?} \label{sec:evolution:methods:clustering}

Given succinct vector representations of subreddit states, we now need to
quantify the distance between these representations. A simple approach is to
rely on euclidean distances between the vectors of two subreddit states,
however, there are two main drawbacks with this approach: (1) it is not clear
what a euclidean distance threshold must be for us to claim that subreddit
states have converged and (2) the euclidean distance metric is not suitable
for measuring the \emph{nature} of subreddit evolution -- \ie the direction
(relative to other communities) of evolution in terms of topics and active user
bases. We overcome these limitations by relying on \emph{agglomerative cluster
distance} as a measure of distance between subreddit states. 

\begin{figure}[t]
\includegraphics[width=\linewidth]{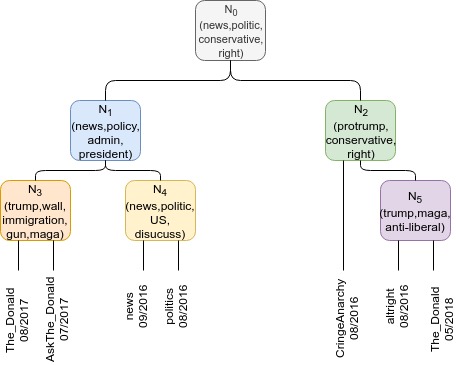}
\caption{An snippet of a larger dendrogram obtained by hierarchical clustering
  on topic vectors. LDA topics are associated with each cluster.}
\label{fig:evolution:methods:dendro}
\end{figure}

\para{Creating subreddit state clusters.} Using the topic and active
    user vectors generated in our previous step, we generated two clustering
    models -- one which clustered subreddit states by their topic vectors and
    another which clustered subreddit states by their active user vectors. In
    both cases, we used agglomerative (bottom-up) hierarchical ward-method
    clustering \cite{murtagh2014ward}. Agglomerative clustering starts off by
    treating each subreddit state individually and merging closest clusters
    together in a step by step process. Typically visualized as a dendrogram,
    clusters at higher levels are larger and more general while clusters at
    lower levels are much smaller and specific. Agglomerative clustering, in
    addition to providing an intuitive notion of inter-cluster distance
    (described below), also provides the ability to cluster without
    {pre-specifying} number of clusters.

  \para{Quantifying magnitude of evolution.} Given our hierarchical
    cluster models based on topic and active user vectors, we used an intuitive
    measure of similarity between two subreddit states. We quantified the
    distance between subreddit states $s_i$ and $s_j$ as the \emph{height} of
    the nearest common ancestor of $s_i$ and $s_j$. To compute this distance,
    we performed a search for the nearest common ancestor of the input subreddit
    states, computed the $\max$ of the number of nodes between this parent and
    each of the subreddit states being compared (\ie $s_i$ and $s_j$).
    Therefore, if the distance between the subreddit states associated with two
    consecutive months of a subreddit is high, it is indicative that the (topic
    or active user) vectors associated with the newer subreddit states are much
    more similar to other subreddits than its own previous state. This is
    indicative of topical or user base evolution towards other subreddits. We
    expect that the evolutionary distance between consecutive months of
    a subreddit will be one or two (\ie nearest parent is at distance one or
    two from each state) in the event of convergence.

  \para{Measuring nature of evolution.} In addition to quantifying the
    magnitude of evolution, we also present a method to measure the nature of
    change between consecutive subreddit states. For this, we relied on Latent
    Dirichlet Allocation (LDA) to assign topics associated with each cluster in
    our hierarchical cluster models. For each subreddit state, we obtained the
    text of its subreddit wiki. This text was then associated with all parents
    of this subreddit state in our hierarchical model. LDA was then performed
    on the text associated with every possible cluster in our model, therefore
    giving us a set of topics associated with every cluster. We also repeated
    this process with the comments associated with each cluster. These topics
    give us an insight into the topical direction in which subreddits are
    evolving. 
    
\Cref{fig:evolution:methods:dendro} shows a snippet of the dendrogram
associated with our topic vector based hierarchical cluster model. As leaves in
the dendrogram we have our subreddit states and with each parent in the tree
are topics extracted from the child subreddit wikis and comments using LDA.
This snippet shows how \subreddit{The\_Donald} moved from being most closely
associated with \subreddit{AskThe\_Donald} in late 2017 to being associated
most closely with the 2016-versions of now banned subreddits --
\subreddit{altright} and \subreddit{CringeAnarchy} in mid-2018. In this
example, we see that \subreddit{The\_Donald} has evolved a distance of
3 between 08/2017 and 05/2018.

%
\subsection{Results}\label{sec:evolution:results}

\begin{figure*}[th]
    \centering
  \begin{subfigure}{.235\textwidth} 
        \includegraphics[width=\textwidth]{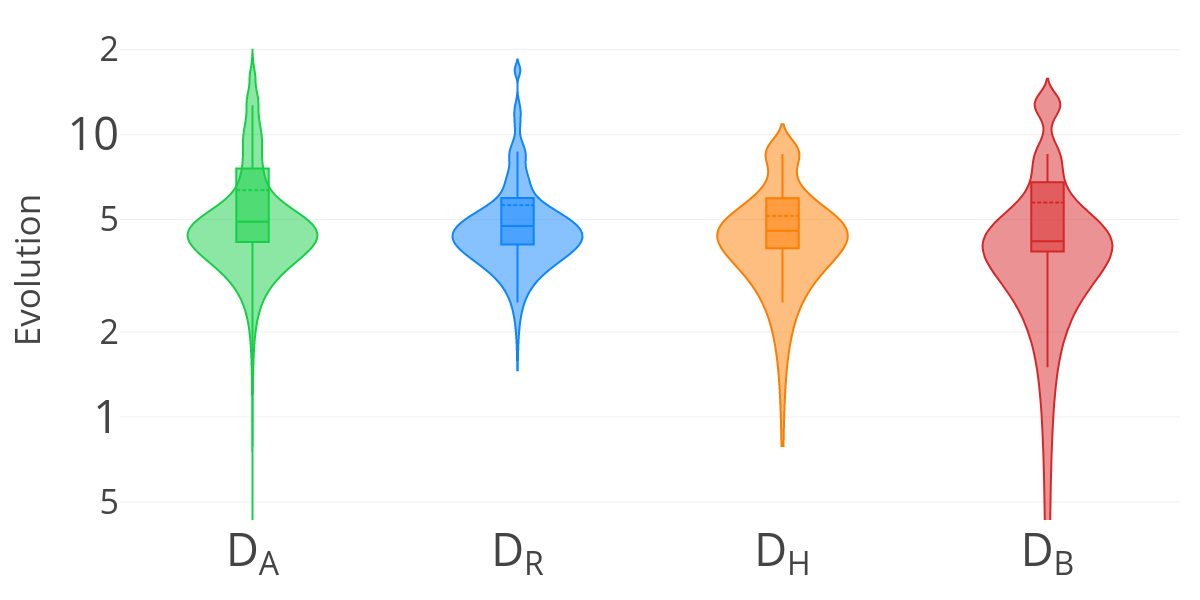}
    \caption{(log scale) Distribution of mean monthly topic evolution.} 
    \label{fig:evolution:results:convergetopic:distribution}
    \end{subfigure}
    \begin{subfigure}{.235\textwidth} 
        \includegraphics[width=\textwidth]{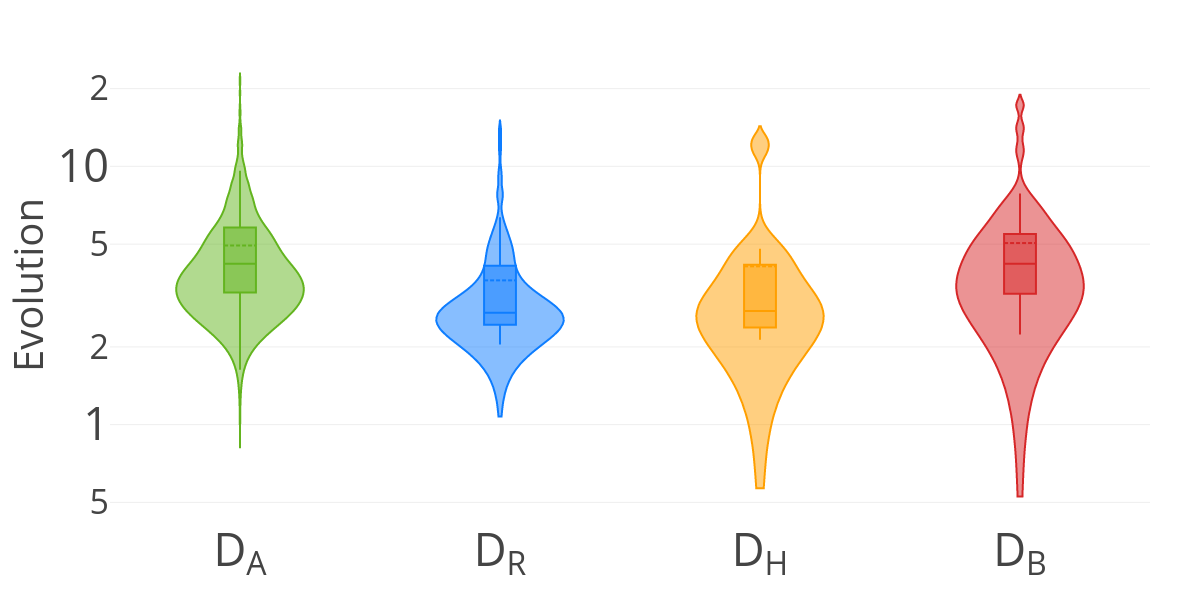}
    \caption{(log scale) Distribution of mean monthly user base evolution.} 
    \label{fig:evolution:results:convergeuser:distribution}
    \end{subfigure}
    \begin{subfigure}{.235\textwidth} 
        \includegraphics[width=\textwidth]{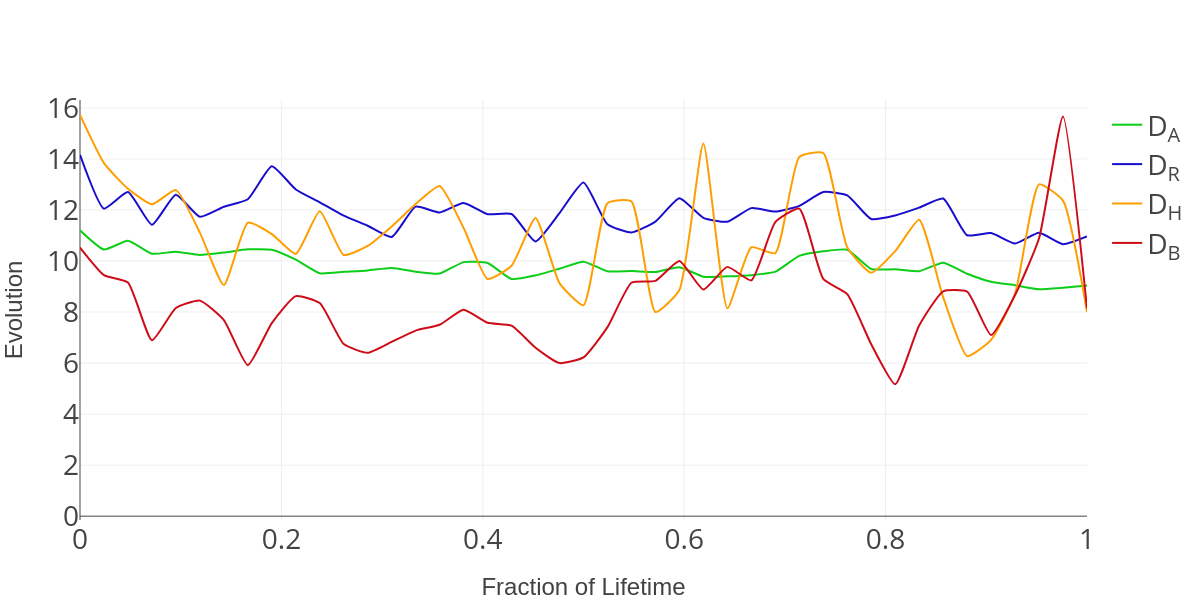}
    \caption{Subreddit age vs magnitude of topic evolution.}
        \label{fig:evolution:results:convergetopic:lifetime}
    \end{subfigure}
    \begin{subfigure}{.235\textwidth} 
        \includegraphics[width=\textwidth]{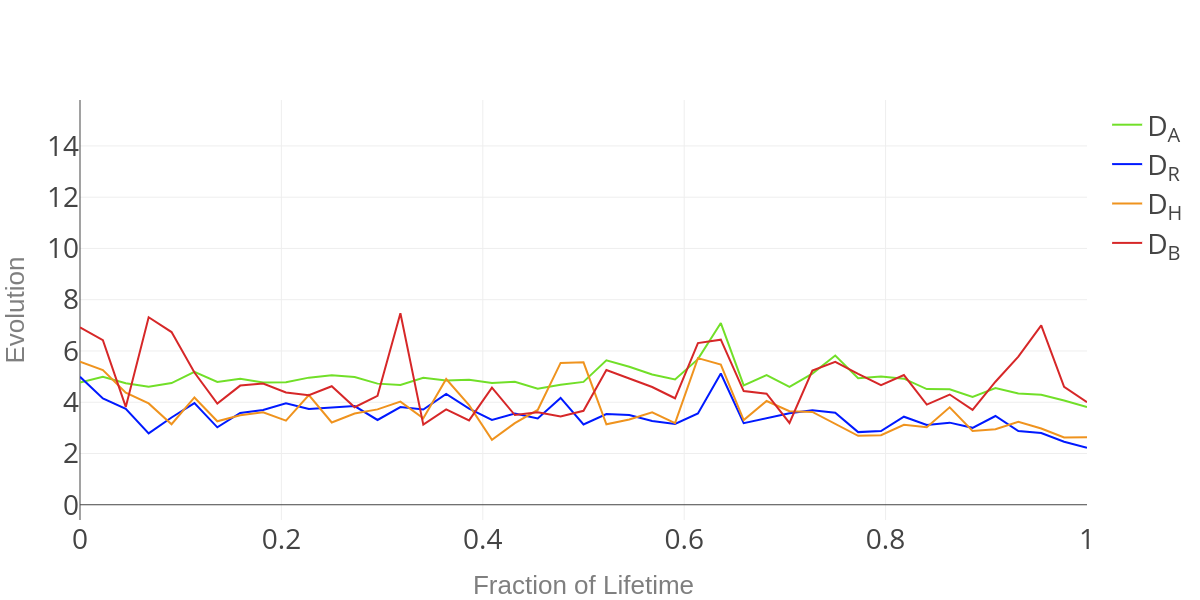}
    \caption{Subreddit age vs magnitude of active user base evolution.}
        \label{fig:evolution:results:convergeuser:lifetime}
    \end{subfigure}

    \caption{Characteristics of topic and active user base evolution for
  subreddits in $\mathfrak{D}_A$, $\mathfrak{D}_B$, $\mathfrak{D}_H$, and
  $\mathfrak{D}_R$. 
  Figures (a) and (b) show the distribution of the mean magnitudes of
  topic and active user base evolution per month for subreddits in different
  categories. Figures (c) and (d) show the average magnitude of topic and
  active user base evolution as a function of subreddit age. The maximum lifetime
  of a subreddit is the time between the first post and last post in our
  datasets (our data collection ended on 10/2018). The `fraction of lifetime'
  is the fraction of this maximum lifetime. }
  \label{fig:evolution:results}
\end{figure*}

We now quantify the magnitude of topical and active user base evolution for
subreddits belonging to different datasets -- \ie $\mathfrak{D}_A$,
$\mathfrak{D}_B$, $\mathfrak{D}_H$, and $\mathfrak{D}_R$. We specifically focus
on measuring the average magnitude of evolution per month for different
subreddits (\Cref{sec:evolution:results:distributions}) and how subreddits
evolve as a function of their age (\Cref{sec:evolution:results:age}).
{Based on these analyses, we expect to be able to verify the hypothesis that
subreddits do not always converge to user or topical stability
(\Cref{sec:evolution:results:takeaway})}. Our results are illustrated in
\Cref{fig:evolution:results}.

\subsubsection{How much do subreddit topics and active user bases evolve per
month?} \label{sec:evolution:results:distributions}

We make the following observations from
\Cref{fig:evolution:results:convergetopic:distribution} and
\Cref{fig:evolution:results:convergeuser:distribution} which shows the
distributions of mean monthly magnitude of topical and active user base
evolution for subreddits in different categories.

  \para{Subreddits have a high average rate of evolution per month.}
    Looking at the average magnitude of topical and active user base
    evolution per month for different subreddit categories, we see that the
    mean values for subreddits in different categories are between 4 and 10.
    This is indicative that the average subreddit evolves at a high rate on
    average. For reference, an average value of 1 or 2 would indicate that the
    subreddit has not evolved since its conception. We also notice that, in
    general, the magnitude of evolution of user bases is lower than of topics.

  \para{Average magnitude of evolution per month is consistent across
    subreddit categories, but the long tails vary.} We see that subreddit
    categories are not a discriminator when considering only the average
    monthly magnitudes of evolution. However, we see that the distribution of
    these magnitudes varies by subreddit category. In particular, we see that
    subreddits in $\mathfrak{D}_B$ and $\mathfrak{D}_H$ are heavier weighted
    towards lower and higher average monthly magnitudes of evolution -- \ie
    subreddits in these groups are more likely to have extremely high or low
    average magnitudes of evolution. 
    This indicates that (1) some subreddits in $\mathfrak{D}_B$ and
    $\mathfrak{D}_H$ show more signs of being likely to converge to topical and
    user base stability, but this is not a common case for any subreddit
    category and (2) different categories of subreddits might have different
    evolutionary patterns -- a promising insight for hypothesis H2 which aims
    to predict the subreddit category based on evolutionary patterns.

In addition to measuring a subreddit's average magnitude of evolution, we are
also able to check the nature of subreddit evolution by observing the keywords
of different subreddit states. We examine the \emph{most evolved} subreddit in each category
to understand the nature of evolution in each case. There are highlighted in
\Cref{tab:evolution:results:special}. Of particular interest are
\subreddit{enoughsandersspam} and \subreddit{altright}. Studying the nature of
evolution of \subreddit{enoughsandersspam} we see that the subreddit initially
started as a subreddit to mock and harass supporters of US politician Bernie
Sanders going as far as goading them into suicide and threatening rape
\cite{ESS-threats}. This behavior resulted in its addition into
$\mathfrak{D}_H$. However, we observe that the subreddit turned into
a pro-Bernie Sanders community in late 2018 as a result of a moderator
takeover \cite{ESS-takeover}. This resulted in an unusually high magnitude of
active user base evolution. Similarly, we see that \subreddit{altright} emerged
as a political movement in early 2016 and evolved into a white supremacist and
anti-Semitic subreddit by early 2017. Our method identifies these
subreddits as the most evolved and is able to hint at the nature of evolution, which serves as validation for the techniques and metrics in
\Cref{sec:evolution:methods}. 

\begin{table}
  \centering
  \small
  \begin{tabular}{p{1in}p{.25in}p{.25in}p{1.25in}}
    \toprule
    \textbf{Subreddit} & \textbf{AMM} & \textbf{LM} & \textbf{Nature of evolution}\\
    \midrule
    ($\mathfrak{D}_B$) \subreddit{altright}           & T:13.7 U:11.5 & T:31 U:17 & \red{right}, \red{support}, \red{vote}, \blue{jew}, \blue{white}, \blue{support} \\
      \midrule
    ($\mathfrak{D}_H$) \subreddit{enoughsandersspam}  & T:10.1 U:12.7 & T:50 U:53 & \red{bernie}, \red{the\_donald}, \red{spam}, \blue{hero}, \blue{obama}, \blue{original} \\
     \midrule
    ($\mathfrak{D}_R$) \subreddit{the\_meltdown}      & T:14.4 U:12.3 & T:31 U:13 &  \red{prison}, \red{wikileaks}, \red{revolution},  \blue{DNC}, \blue{propaganda} \\
     \midrule
    ($\mathfrak{D}_H$) \subreddit{GalaxyNote7}        & T:30.7 U:22.0 & T:50 U:88 & \red{samsung}, \red{battery}, \red{nexus}, \blue{deals}, \blue{free}, \blue{blackfriday} \\
    \bottomrule
  \end{tabular}
  \caption{Subreddits with highest monthly evolution magnitudes in each
  category and the nature of their evolution. \textbf{AMM} is the average
  magnitude of evolution per month, \textbf{LM} is the magnitude of evolution
  over the entire lifetime of the subreddit, T and U denote the magnitudes of
  topic and active user base evolution, \red{Red} keywords represent
  topics associated with the start of the subreddit's lifetime, and \blue{Blue}
  keywords represent topics associated with the end of the subreddit's
  lifetime which is characterized by when the subreddit was banned or the end
  of our data collection (10/2018).}
  \label{tab:evolution:results:special}
\end{table}

\subsubsection{How do subreddit topics and active user bases evolve as
a function of subreddit age?} \label{sec:evolution:results:age}

Our previous result shows that there is, on average, a significant magnitude of
evolution per month for subreddits in all our categories. However, owing to the
long tail distributions, we are unable to conclude that subreddits do not
converge to topics or user bases. We now breakdown the magnitude of evolution
by subreddit age. In the event of convergence, we expect to see that the
magnitude of evolution drops to and stays at the minimum (1 or 2) after
a certain point in the lifetime of a subreddit.
\Cref{fig:evolution:results:convergetopic:lifetime} and
\Cref{fig:evolution:results:convergeuser:lifetime} show how subreddit topics 
and active user bases evolve as a function of their age. We can make the
following observation from these plots. 

  \para{Banned subreddits ($\mathfrak{D}_B$) show high magnitude
    evolutions before being banned.} 
    \Cref{fig:evolution:results:convergetopic:lifetime} shows that the
    magnitude of topical evolution per month is generally high for all
    subreddit categories, regardless of age of the subreddit. However, across
    different categories we observe a difference. Particularly in the case of
    $\mathfrak{D}_B$ we see that the magnitude of evolution is significantly
    lower for most of the subreddit's lifetime and a sharp increase occurs
    right before the subreddit is banned. We see a similar rapid increase in
    magnitude even when considering user base evolution (shown in
    \Cref{fig:evolution:results:convergeuser:lifetime}). This is indicative of
    a rapid change in topic and community right before a ban event --
    unfortunately, a causal relationship cannot be inferred from our data.
We also analyze the average net evolution over a lifetime for subreddits in
each category. Here again we find that subreddits in $\mathfrak{D}_B$ and
$\mathfrak{D}_H$ have significantly higher average magnitudes of topic
evolution (14.7 and 18.4, respectively) than other subreddits
($\mathfrak{D}_A$ has a mean topic evolution magnitude of 8.2). This is
indicative that subreddits which get banned or get labeled as hateful often do
not start with discussions or user bases which occur at the end of their
lifespan.

\subsubsection{Takeaway: Do subreddits converge to topical or user stability?}
\label{sec:evolution:results:takeaway}

Taken together, our results show that subreddits generally have a high average
magnitude of topic and user base evolution (per month and over a lifetime) and
this evolution, in most cases, is independent of the age of the subreddit.
Therefore, \emph{we are unable to claim that subreddits always converge to
topical or user stability -- \ie hypothesis H1 is valid}. This suggests that
policies that involve verifying and applying moderation policies at the time of
subreddit creation is not sufficient. Rather, the application of moderation
policies need to be considered repeatedly over the course of the subreddit's
lifetime. This suggests a very high attention cost for human administrators
and moderators who need to (1) monitor how thousands of subreddits are evolving
and (2) identify how and when to act to maintain community civility.

\section{Predictors of Hateful or Dangerous Subreddits}\label{sec:predictors}

In addition to showing that human administrators and moderators seeking to
maintain civility need to constantly monitor communities for changing topics
and user bases, our previous results (\Cref{sec:evolution:results}) show that 
banned and hateful subreddits have different evolutionary characteristics than
other subreddits. Specifically, we showed that (1) subreddits in these groups
were more likely to have smaller average monthly evolution magnitudes --
particularly when considering topic evolution, (2) have larger magnitudes of
evolution over a lifetime, and (3) have a few high magnitude evolution events
-- typically in the latter half of the subreddit lifespan. All these results
suggest that it might be possible to identify subreddits likely to evolve into
hateful or dangerous subreddits based on measurable evolutionary
characteristics. In this section, we use the above insights to test the
following hypothesis: \textbf{H2. Evolution into hateful or dangerous
subreddits can be predicted}. This hypothesis, if valid will show that tools
may be built to help moderators pre-emptively identify and watch subreddits
likely to devolve into offensive or dangerous subreddits. If invalid, the
failed hypothesis will show that evolutionary characteristics cannot be used to
motivate pre-emptive moderation actions to mitigate the impact of hateful or
dangerous subreddits.
Our methodological and result contributions are outlined
in \Cref{tab:predictors}.

\begin{table}
  \centering
  \small
  \begin{tabular}{p{2.7in}l}
    \toprule
    \textbf{Methodological questions} & \textbf{\Cref{sec:predictors:methods}}\\
    \midrule
    What are hateful or dangerous subreddits?
    & \Cref{sec:predictors:methods:labels}\\
    What evolutionary features do we extract for each subreddit?
    & \Cref{sec:predictors:methods:features}\\
    How do we identify the predictive values of features?
    & \Cref{sec:predictors:methods:weights}\\
    \midrule
    \textbf{Research questions} & \textbf{\Cref{sec:predictors:results}}\\
    \midrule
    Can we identify hateful or dangerous subreddits by their evolutionary
    features? & \Cref{sec:predictors:results:accuracy}\\
    What features are most important for predicting the evolution into hateful
    or dangerous subreddits? & \Cref{sec:predictors:results:features}\\
    How far ahead of time can hateful or dangerous subreddits be predicted?
    & \Cref{sec:predictors:results:early}\\
    Can evolution into hateful or dangerous subreddits be predicted?
    & \Cref{sec:predictors:results:takeaway}\\
    \bottomrule
  \end{tabular}
  \caption{Hypothesis 2: Evolution into hateful or dangerous subreddits can
  be predicted. Summary of methods and results.}
  \label{tab:predictors}
\end{table}

\para{High-level overview.} Our goal is to identify
the features that are good predictors of future subreddit behavior.
Specifically, we wish to identify the features that can be
useful to predict the likelihood of a subreddit being classified as hateful or
dangerous in the future. To achieve our goal, we first need to understand what
subreddits are \emph{hateful} or \emph{dangerous}. Once we have methods to
assign these labels, we need a method to identify the evolutionary
characteristics of subreddits which are predictors of subreddits being
classified as hateful or dangerous. We accomplish this using the following
approach, which is explained in more detail in \Cref{sec:predictors:methods}.
First, we assign \emph{hate} labels to all subreddits in our $\mathfrak{D}_H$
dataset, \emph{dangerous} labels to all subreddits in our $\mathfrak{D}_B$
dataset, and \emph{other} labels to all subreddits in our $\mathfrak{D}_A$
dataset. Next, we break the lifespan of each subreddit down into four quarters
and extract features from each of these periods. Our features come from four
categories -- community-related, moderator-related, user-related, and
structure-related. 

Once we have these features and labels for each subreddit,
we fit several explainable classification models including a logistic
regression, decision tree, and random forest. 
We argue that if our explainable classifiers have reasonably high accuracy in
predicting the labels assigned to each subreddit, then the features that are
highly weighted by them \emph{must be good predictors for hateful and
dangerous subreddits}. We present our results (\Cref{sec:predictors:results}),
which broadly validate our proposed hypothesis by demonstrating that (1) we
can achieve high classification accuracy with relatively simple evolutionary
features and (2) hateful and dangerous subreddits can be identified with
reasonable accuracy very early in their lifetime. 

\subsection{Methods}\label{sec:predictors:methods}

Our method for testing the proposed hypothesis requires us three steps. First,
we need a method to assign {hateful}, {dangerous}, and {other} labels to
subreddits in our dataset (\Cref{sec:predictors:methods:labels}). Next, we need
to identify different features to extract from our subreddits
(\Cref{sec:predictors:methods:features}). Finally, we evaluate the predictive
value of these features (\Cref{sec:predictors:methods:weights}). 

\subsubsection{What are hateful or dangerous subreddits?}
\label{sec:predictors:methods:labels}

We rely on our previously described (in \Cref{sec:background:data}) datasets of
banned and quarantined subreddits ($\mathfrak{D}_B$), hateful subreddits
($\mathfrak{D}_H$), and a random sample of 190 of the most active subreddits
($\mathfrak{D}_A$) as sources of dangerous, hateful, and other subreddits,
respectively. This labeling is acceptable since subreddits from
$\mathfrak{D}_B$ were banned or quarantined for violating one of Reddit's
content policies -- a clearly unacceptable and dangerous behavior. Therefore,
by analyzing the features most likely to be predictive for subreddits in
$\mathfrak{D}_B$, we are able to identify evolutionary characteristics which
are likely to lead to violations of Reddit's content policies. Similarly,
subreddits from $\mathfrak{D}_H$ were found to be the most frequently reported
non-banned but hateful subreddit by the wider reddit community
(\subreddit{againsthatesubreddits} and \subreddit{SubredditDrama}) and the
media for promoting sexist and racist hate speech. Therefore, by analyzing the
features most likely to be predictive for subreddits in $\mathfrak{D}_H$, we
are able to identify evolutionary characteristics which are likely to lead to
large number of community complaints. Predictive features for both
$\mathfrak{D}_B$ and $\mathfrak{D}_H$ are useful for sitewide moderators and
administrators to narrow down the likely causes of future content policy
violations and community complaints -- enabling more effective monitoring or
pre-emptive moderation interventions.

\subsubsection{What evolutionary features do we extract for each subreddit?}
\label{sec:predictors:methods:features}

For each subreddit in our datasets, we break their lifespan into four quarters
and extract features from each of these quarters. This allows us to do several
things: (1) we are able to get an identical number of features from all
subreddits -- even if they have vastly different lifespan values, and (2) we
are able to capture features from different phases in the evolution of
a subreddit. Our extracted features fall in four categories --
community-, moderator-, user-, and structure-related features. These were
largely influenced by existing literature seeking to predict community
dynamics (\Cref{sec:related:evolution} describes and compares this work and its
influence on ours). All extracted features are shown in
\Cref{tab:predictors:features}.

  \para{Community-related features.} This category of features captures
    the dynamics of the interactions occurring within the community -- \eg how
    large is the active community, how highly do community members rate each
    others posts, \etc

  \para{Moderator-related features.} Moderators play a large rule in
    directing the growth and policies within each community. This category of
    features captures how the moderator team interacts with the community --
    \eg how many posts or comments are deleted, how frequently do moderators
    create stickied posts, \etc 

  \para{User-related features.} This category of features captures
    characteristics of the average users within the community -- \eg how active
    are users, how frequently do they delete their posts?

  \para{Structure-related features.} Finally, we also introduce
    a category of features to capture how a subreddit is connected (in terms of
    shared user base) to other communities -- \eg how isolated is the
    subreddit, what fraction of its connections are to other communities which
    were previously classified as hateful or banned, \etc

\begin{table}[t]
\small
  \begin{tabular}{p{.55in}ll}
  \toprule
    {\textbf{Category}}  & {\textbf{Features}} & {\textbf{Type}} \\ \hline
                                          & \# active unique posters                  & Point      \\ 
                                          & \# active unique commenters               & Point      \\
                                          & \# posts                                  & Point      \\
                                          & \# comments                               & Point      \\
                                          & Dist. of comments \& posts                & Quartiles  \\
                                          & Dist. of score \& posts                   & Quartiles  \\
                                          & Dist. of score \& comments                & Quartiles  \\
                                          & \% of active user growth                  & Point      \\
                                          & Dist. of controversial score per post     & Quartiles  \\
                                          & \# controversial comments                 & Point      \\ 
                                          & \# gilded posts                           & Point      \\
    \multirow{-12}{*}{\textbf{Community}} & \# gilded comments                        & Point      \\ 
    \midrule
                                          & \# moderators                             & Point      \\
                                          & \# stickied posts                         & Point      \\
                                          & \# removed comments                       & Point      \\
    \multirow{-4}{*}{\textbf{Moderators}} & \# removed posts                          & Point      \\
    \midrule
                                          & \# active months                          & Quartiles  \\
    \multirow{-2}{*}{\textbf{Users}}      & \# deleted comments                       & Quartiles  \\
    \midrule
                                          & \# uniquely connected communities         & Point      \\
                                          & \# total connections                      & Point      \\
                                          & \# total connections to banned communities  & Point    \\ 
                                          & \# total connections to hateful communities & Point    \\
                                          & \% of connections to banned communities     & Point    \\
    \multirow{-6}{*}{\textbf{Structural}} & \% of connections to hateful communities    & Point    \\
  \bottomrule
\end{tabular}
 \caption{List of features extracted from each quarter of a subreddit's
  lifespan.}
 \label{tab:predictors:features}
\end{table}

\subsubsection{How do we identify the predictive value of features?}
\label{sec:predictors:methods:weights}

Given labels for each subreddit and a set of features associated with each
stage in its lifetime, we now seek to understand the predictive values of these
features. We achieve this in two steps: First, we build a machine learning
classifier which uses these features to predict the labels associated with
each subreddit. Next, we analyze the weights associated with each feature by
the classifier. Our argument is that \emph{if a classifier is able to achieve
a reasonably high accuracy, then the features it weighs heavily must have some
predictive value}. We focus solely on interpretable classifier models (logistic
regressions, decision trees, and random forests) since we need to obtain
feature weights. However, we see very similar classifier performance, when
comparing our models to other non-interpretable models such as SVMs and neural
networks. To evaluate the performance of each classifier we perform 10-fold
cross-validation and report the accuracy and F1-scores. 
Once we have fitted and cross-validated a classifier model, we interpret them
as follows.

  \para{Logistic regression model.} {Logistic Regression} models
    a relationship between an outcome variable $y$ and a group of predictor
    variables in terms of log odds. In order to interpret the model, we compute
    the estimated weights for each feature and their corresponding odds
    ratio \cite{ML-Interpret}. If the odds ratio for a feature ($f$) is $x$, it
    means that a unit increase in $f$ changes the odds of our outcome variable
    $y$ by a factor of $x$ when all other features remain the same. By
    calculating the features with the highest odds ratios for different labels,
    we are able to identify which features are the best predictors of dangerous/banned,
    hateful, and other/benign subreddits.

  \para{Decision tree and random forest models.} Tree based models are
    the easiest to interpret. We find the importance of each feature using Gini
    Importance \cite{Breiman-Gini}. At a high-level the gini importance counts
    the number of times a feature is used as a splitting variable, in
    proportion with the fraction of samples it splits. For random forests, the
    gini importance is averaged over all the constructed trees. We expect more important features to have higher gini importance scores. Unlike
    logistic regression interpretation, a limitation here is that this metric
    only allows us to rank feature importance, but not quantify the relative
    difference of their importance. Further, we do not observe per-outcome
    importances.

\subsection{Results}\label{sec:predictors:results}

We now focus on measuring the effectiveness of using evolutionary features to
predict which category (from dangerous/banned, hateful, and other) subreddits
belong to (\Cref{sec:predictors:results:accuracy}), which features are the best
predictors of subreddit behavior (\Cref{sec:predictors:results:features}), and
the impact of observation time on classifier accuracy
(\Cref{sec:predictors:results:early}). Our results are summarized in 
\Cref{tab:predictors:results}.

\begin{table*}[t]
  \small
  \centering
  \begin{tabular}{l X{.5in} X{.5in} X{.65in} X{.5in} p{2.75in}}
    \toprule
    \textbf{Model} & \multicolumn{4}{c}{\textbf{Classifier Performance
    (\%Accuracy, \%F1)}} & \textbf{Top features} \\\cmidrule{2-5}
                   &  Q1 & Q1 + Q2 & Q1 + Q2 + Q3 & Total & \\\midrule
    Logistic regression  & (74, 48) & (77, 54) & (78, 61) & (84, 72)
    & \%connections to banned and hateful communities, 
      \#stickied posts and \#moderators. \\
    Decision tree        & (82, 70) & (85, 74) & (83, 69) & (83, 71) 
    & \#total connections to hateful and banned communities, 
      distribution of post scores. \\
    Random forest        & (89, 82) & (90, 83) & (90, 83) & (90, 84) 
    & \#total connections to hateful and banned communities, 
      \%connections to banned and hateful communities. \\
    \bottomrule
  \end{tabular}
  \caption{Performance of our classifier models and the most important class of
  features used by the models. Q1 denotes that the classifier only had access
  to features from the first quarter of the subreddit lifespans, Q1+Q2 denotes
  access to the first half of the lifespan, Q1+Q2+Q3 denotes access to the
  first three-quarters, and Total denotes access to the entire lifespan.}
  \label{tab:predictors:results}
\end{table*}

\subsubsection{Can we identify hateful or dangerous subreddits by their
evolutionary features?} \label{sec:predictors:results:accuracy}

Column ``Total'' of \Cref{tab:predictors:results} shows how our different
explainable classifier models performed at classifying subreddits into banned
(or, dangerous), hateful, and other when given access to all evolutionary
features of the subreddit. As we can see all our models perform reasonably well
achieving F1-scores from 71\% (decision tree) to 84\% (random forests)
and accuracies of 83\% (decision tree) to 90\% (random forests) --
compared to a baseline accuracy of 58\% (for a classifier which simply outputs
the most dominant label in our dataset -- \ie `other'). This promising result
shows that evolutionary features can be used as a distinguisher between each
class of subreddit. Further, in observing the confusion matrices obtained
from our testing, we see that most classifiers had trouble distinguishing
between subreddits with `dangerous/banned' and `hate' labels. In particular,
several hate subreddits (\eg \subreddit{The\_Donald} and
\subreddit{inceltears}) were consistently misclassified as dangerous/banned
subreddits due to very similar evolutionary patterns. We note that since our
classes rely on human coding (\ie our banned/dangerous subreddits are
essentially labeled by Reddit administrators), it is not clear if these
misclassifications are a result of inadequate features or inconsistent
application of content policies by Reddit administrators (although a cursory
qualitative analysis of the content in \subreddit{The\_Donald} suggests the
latter). \emph{The high accuracy of our models suggest that interpreting their
use of evolutionary features will yield predictors of subreddit behaviors}.

\subsubsection{What features are most important for predicting the evolution
into hateful or dangerous subreddits?} \label{sec:predictors:results:features}

Column ``Top features'' of \Cref{tab:predictors:results} shows the most
important features for each of our classifier models using the interpretations
described in \Cref{sec:predictors:methods:weights}. Here we see a common theme
emerge: \emph{structural features capturing the connectivity of subreddit
members to other hateful and dangerous/banned subreddits is the most
discriminative predictor of subreddit behavior}. This feature appears as our
highest ranked feature for all classification models. Interpreting the logistic
regression, we see the feature has an odds ratio of 3.96 -- indicating very
strong predictive value. Our findings here suggest that past participation of
current community members in hateful and banned subreddits is the strongest
predictor of future community behavior. Other features of importance include
moderator team size and activity (measured by \#stickied posts).

\subsubsection{How far ahead of time can hateful or dangerous subreddits be
predicted?} \label{sec:predictors:results:early}

Our goal goes beyond identifying the labels associated with a subreddit. Since
we are motivated by the need for tools to aid proactive (and early) community
interventions, we also need to understand how early the future behavior of
a subreddit can be predicted. To this end, we evaluated each classifier's
performance using features extracted only from the first quarter (Q1), first
half (Q1+Q2), and first three-quarters (Q1+Q2+Q3) of a subreddit's lifespan.
For reference, subreddits in $\mathfrak{D}_B$, which we label as dangerous, had
the following lifespan distribution [$\min$: 7 months, $25^{th}$ percentile: 20
months, median: 34 months, $75^{th}$ percentile: 39 months, $\max$: 43
months]\footnote{These are lower bounds on lifespan of dangerous/banned
subreddits since our data collection only started in 01/2015 and subreddits in
$\mathfrak{D}_B$ might have been active before our data collection began.}. 
From the results shown in \Cref{tab:predictors:results}, we see that while
features from Q4 are the most important for the logistic regression and
decision tree models, the random forest classifier has consistently strong
predictive performance even in the first quarter of the subreddit's lifespan. 
\emph{This shows that even features chosen from early in the subreddits
lifetime (\eg Q1) can be predictors of future subreddit behavior}. This finding
strongly reinforces the feasibility of early and pre-emptive interventions to
correct subreddit behavior.

\subsubsection{Takeaway: Can evolution into hateful or dangerous subreddits be
predicted?} \label{sec:predictors:results:takeaway}

Taken together, our results show that evolutionary features can not only be
used to classify current subreddit behavior, but also predict their future
behavior and that features related to how the community is connected to other
subreddits (\ie network structural features) are the strong predictors.
Therefore, we are able to confirm that \emph{evolution into hateful or
dangerous subreddits can be predicted -- \ie hypothesis H2 is valid}. This
suggests that administrator and community moderation tools which rely on
measuring the connectivity of subreddits to known hate or banned subreddits 
can be used to pre-emptively identify which subreddits require careful
monitoring or even administrator/moderator interventions.

\section{Impact of Moderation}\label{sec:moderation}

So far our results have shown that topics and user bases associated with
subreddits are, on average, evolving at a high rate (\Cref{sec:evolution}) and
the characteristics of this evolution can be used to predict future subreddit
behavior -- particularly evolution into dangerous or hateful subreddits
(\Cref{sec:predictors}). These findings suggest that community administration
and moderation interventions can be used to pre-emptively mitigate the impact
of such communities. In this section, we focus on understanding (1) the impact
of no pre-emptive intervention on known dangerous and hateful communities and
(2) the impact of interventions such as bans and quarantines on known dangerous
and hateful communities. Specifically, we test the hypothesis:
\textbf{H3. (a) User participation in hateful or dangerous subreddits negatively
changes the nature of their participation within the broader community and (b)
this can be corrected by community bans and quarantines}. These hypotheses, if
valid, will show that community bans and quarantines as pre-emptive moderator
actions can reduce offensiveness within the broader community. If
H3(a) is invalid, we will have shown that user behavior in outside communities
is not influenced by their participation in hateful or dangerous communities.
If H3(b) is invalid, we will have shown that there is a need for more nuanced
pre-emptive interventions to mitigate the impact of hateful and dangerous
communities.
 Our methodological and result contributions are highlighted in
\Cref{tab:moderation}.

\begin{table}
  \centering
  \small
  \begin{tabular}{p{2.7in}l}
    \toprule
    \textbf{Methodological questions} & \textbf{\Cref{sec:moderation:methods}}\\
    \midrule
    How do we quantify the nature of user participation?
    & \Cref{sec:moderation:methods:user-behavior}\\
    How do we quantify the impact of events on user participation?
    & \Cref{sec:moderation:methods:interventions}\\
    \midrule
    \textbf{Research questions} & \textbf{\Cref{sec:moderation:results}}\\
    \midrule
    What is the impact of joining a hateful or dangerous community on broader
    user behavior? & \Cref{sec:moderation:results:joining}\\
    What is the impact of banning and quarantining hateful or dangerous
    communities on the behavior of their members?
    & \Cref{sec:moderation:results:intervention}\\ 
    Do ban and quarantine interventions improve user behavior?
    & \Cref{sec:moderation:results:takeaways}\\
    \bottomrule
  \end{tabular}
  \caption{Hypothesis 3: (a) User participation in hateful or dangerous
  subreddits negatively changes the nature of their participation within the
  broader community and (b) this can be corrected by community bans and
  quarantines. Summary of methods and results.}
  \label{tab:moderation}
\end{table}

\para{High-level overview.} At a high-level, our goal is to quantify the change
in user behavior, in the broader community, that occurs as a result of three
types of events: (1) joining a known hateful or dangerous subreddit, (2) the
banning of a community they participate in, and (3) the quarantining of
a community they participate in. To achieve this goal, we need to develop
methods to (1) quantify user behavior and (2) measure the impact of each of the
above events on these metrics. We accomplish this using the following approach,
which is explained in more detail in \Cref{sec:moderation:methods}. First, we
quantify user behavior using two metrics -- incidence of offensive comments and
quantity of participation in known hateful and dangerous communities. Second,
to understand the impact of certain events on their behavior, we carefully
select treatment (users impacted by the event) and control (users similar to
control group but not impacted by the event) groups and analyze how their
behaviour varies before and after the event.

\subsection{Methods} \label{sec:moderation:methods}

Our method for testing the proposed hypothesis requires two main methods.
First, we need to develop metrics to quantify user behavior
(\Cref{sec:moderation:methods:user-behavior}). Next, we need to determine how
to quantify the impact of an event on user behavior
(\Cref{sec:moderation:methods:interventions}).

\subsubsection{How do we quantify the nature of user participation?}
\label{sec:moderation:methods:user-behavior}

We quantify user behavior using two different metrics -- (1) the incidence rate
of offensive comments and (2) the fraction of their participation that occurs
in communities known to be hateful or dangerous. In order to measure the
incidence rate of offensive comments, we use an offensive speech classifier
\cite{hateoffensive} to detect offensive comments made by users. By running
this classifier on each comment made by a user in a month, we are able to
assign an incidence rate of offensiveness for each user for each month. In
addition, for each user for each month, we also record the fraction of user
activity (in terms of comments made) that occurred in dangerous
($\mathfrak{D}_B$) and hateful ($\mathfrak{D}_H$) communities.

\subsubsection{How do we quantify the impact of events on user participation?
} \label{sec:moderation:methods:interventions}

We analyze the impact of three types of events on a user's behavior: (1)
joining a known hateful or dangerous subreddit, (2) the banning of a community
they participate in, and (3) the quarantining of a community they participate
in. By tracking user behavior for the months before and after they were
impacted by one of these events, we are able to see general trends which
illustrate the impact of the event. We go one step further in our analysis of
the impact of each events -- we compare user behavior with a control group
which is constructed as follows. For each user ($u_{t}$) in our treatment group
(\eg user who joined a known hateful or dangerous subreddit), we select the
user ($u_c$) who was not impacted by the corresponding event (\eg did not join
the corresponding hateful subreddit) but shared the most similar community
participation profile as $u_t$. We add $u_c$ to our control group. By comparing
the behavior of users in our control and treatment group, we can make
a stronger case for causal relationships between user behavior and specific
events. 

\subsection{Results} \label{sec:moderation:results}

We now focus on using our described methods (\Cref{sec:moderation:methods}) to
understand the impact of community join
(\Cref{sec:moderation:results:joining}), ban and quarantine 
(\Cref{sec:moderation:results:intervention}) events on user behavior. 

\subsubsection{What is the impact of joining a hateful or dangerous community
on broader user behavior?}\label{sec:moderation:results:joining}

\begin{figure*}[th]
  \centering
  \begin{subfigure}{.235\textwidth}
        \includegraphics[width=\textwidth]{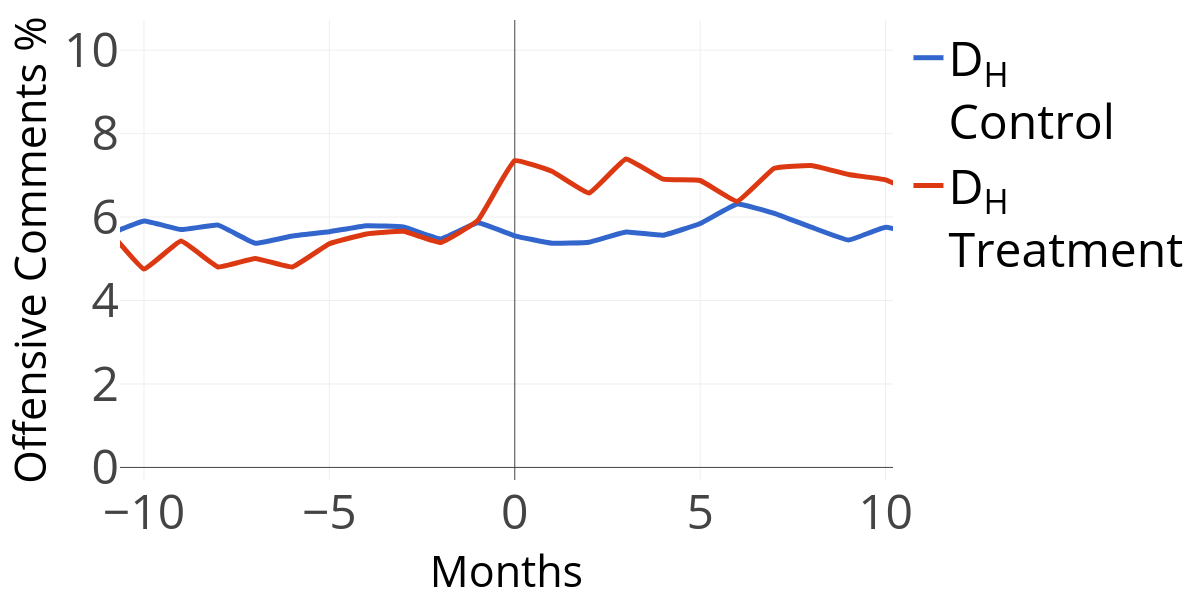}
        \caption{Metric: Offensiveness rate. Event: Joining hateful subreddit
        ($\mathfrak{D}_H$).}
        \label{fig:moderation:results:joining:hateful:offensiveness}
    \end{subfigure}
    \begin{subfigure}{.235\textwidth}
        \includegraphics[width=\textwidth]{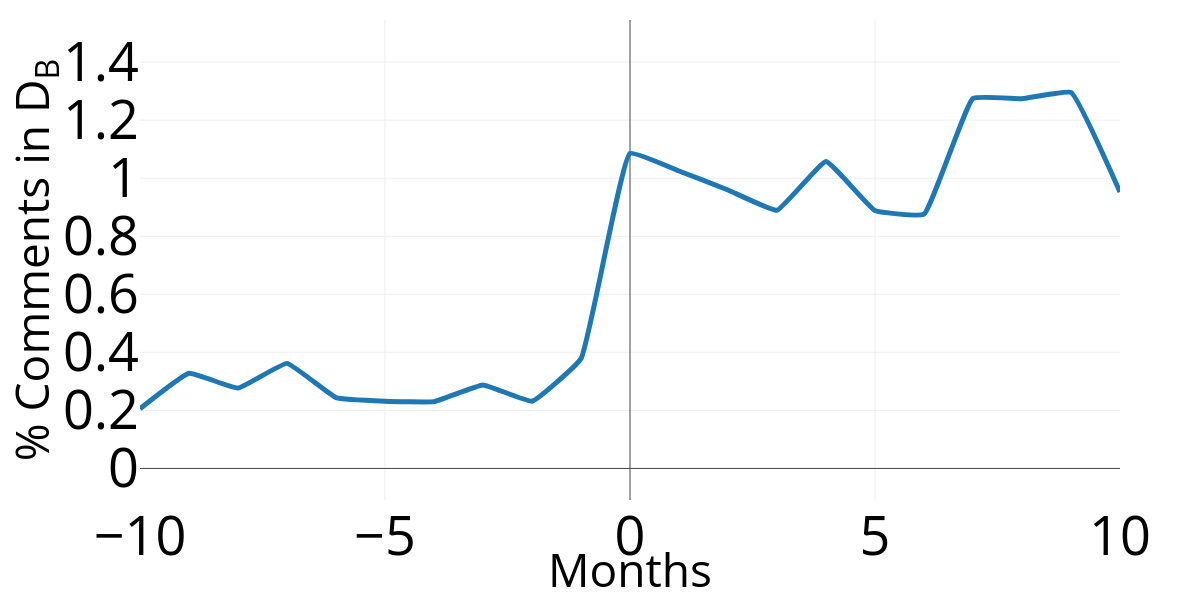}
        \caption{Metric: Rate of participation in dangerous communities
        ($\mathfrak{D}_B$). Event: Joining hateful subreddit ($\mathfrak{D}_H$).}
        \label{fig:moderation:results:joining:hateful:community}
    \end{subfigure}
    \begin{subfigure}{.235\textwidth}
        \includegraphics[width=\textwidth]{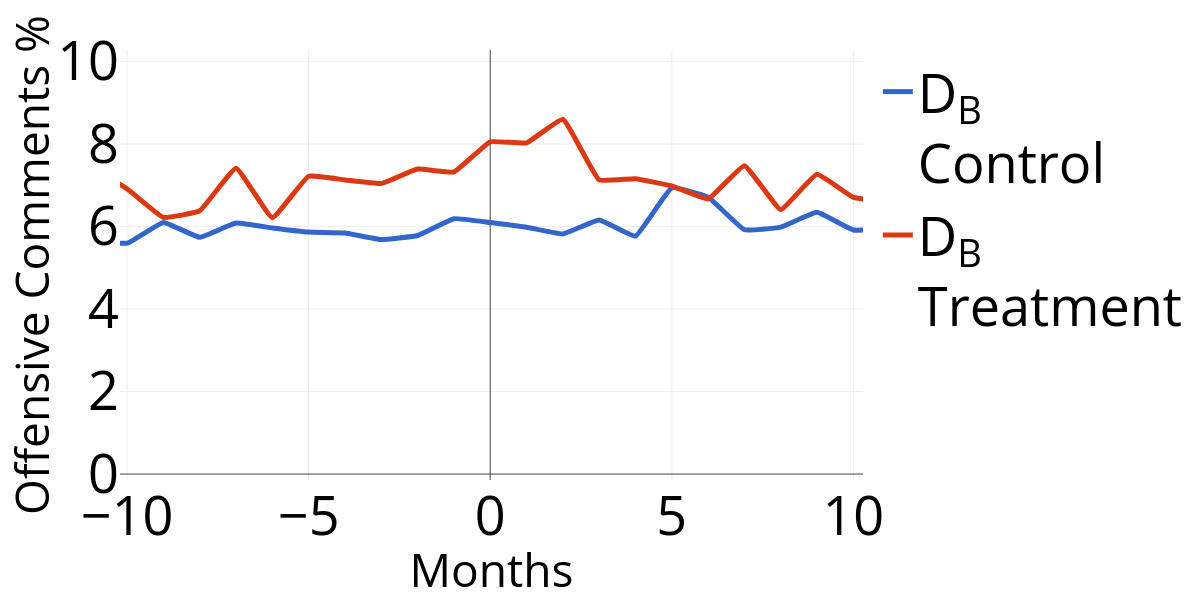}
        \caption{Metric: Offensiveness rate. Event: Joining dangerous subreddit
        ($\mathfrak{D}_B$).}
        \label{fig:moderation:results:joining:banned:offensiveness}
    \end{subfigure}
    \begin{subfigure}{.235\textwidth}
        \includegraphics[width=\textwidth]{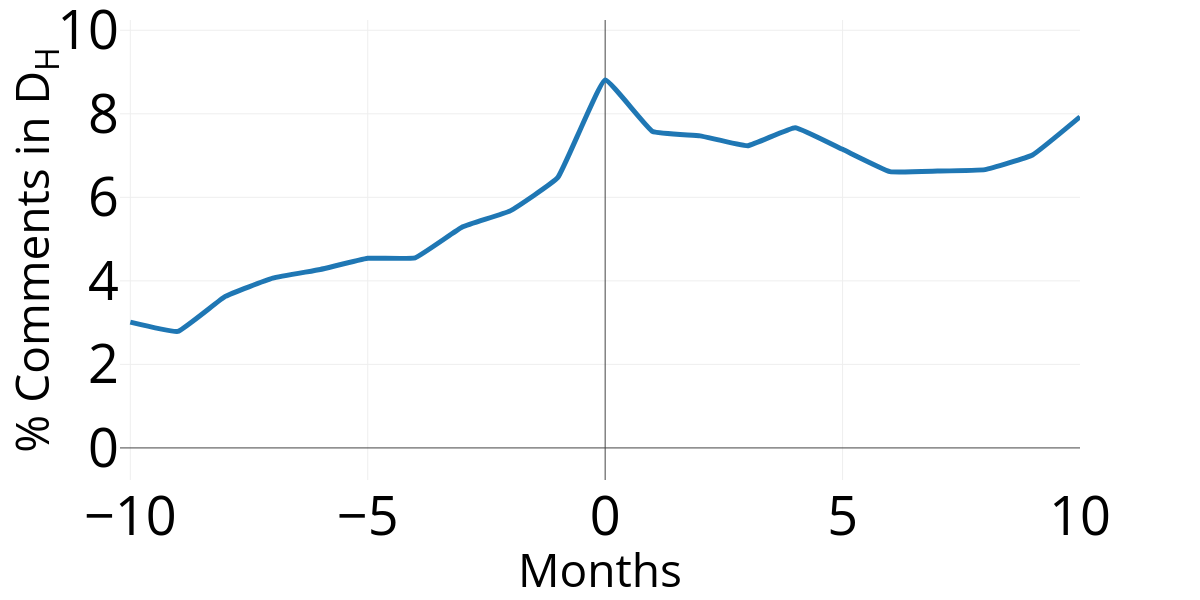}
        \caption{Metric: Rate of participation in hateful communities
        ($\mathfrak{D}_H$). Event: Joining dangerous subreddit ($\mathfrak{D}_B$).}
        \label{fig:moderation:results:joining:banned:community}
    \end{subfigure}
    \caption{The impact of joining events on user behavior. Month ``0''
    represents the month of the join event.}
  \label{fig:moderation:results:joining}
\end{figure*}

\Cref{fig:moderation:results:joining} shows the impact of joining dangerous
($\mathfrak{D}_B$) and hateful ($\mathfrak{D}_H$) communities on user behavior.
We can make several interesting observations from these results. 

\para{Joining hateful communities increases offensiveness in the broader
community.} Looking at
\Cref{fig:moderation:results:joining:hateful:offensiveness} we see that during
the month of the join event associated with hate subreddits, our treatment
users show an increase in offensive behavior. Interestingly, we see that
our treatment users who joined dangerous subreddits show signs of increased
offensiveness from our control group for many months before the join event
(\Cref{fig:moderation:results:joining:banned:offensiveness}). Seeking to better
understand this trend leads us to our next observation.

\para{Hateful communities are a pipeline to dangerous communities.} Looking at
\Cref{fig:moderation:results:joining:hateful:community}, we see that joining
a hate community leads to nearly 500\% more activity in dangerous communities.
Alarmingly, this increase happens within the same month of the join event and
remains high for many months after. This is strong evidence of a causal
relationship between joining hate communities and participation in dangerous
communities. We do not see a similar trend when analyzing the rate of
participation in hateful communities after joining a dangerous community
(\Cref{fig:moderation:results:joining:banned:community}).

\subsubsection{What is the impact of banning and quarantining hateful or
dangerous communities on the behavior of their members?}
\label{sec:moderation:results:intervention}

\begin{figure*}[th]
  \centering
  \begin{subfigure}{.235\textwidth}
        \includegraphics[width=\textwidth]{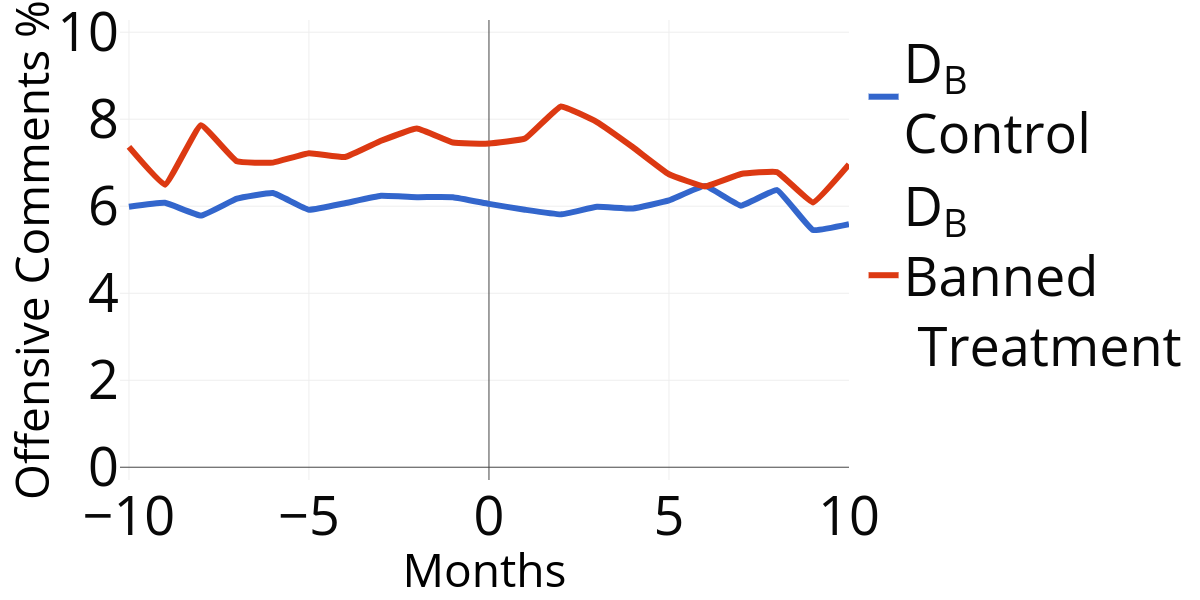}
        \caption{Metric: Offensiveness rate. Event: Banning dangerous subreddit
        ($\mathfrak{D}_B$).}
        \label{fig:moderation:results:banning:offensiveness}
    \end{subfigure}
    \begin{subfigure}{.235\textwidth}
        \includegraphics[width=\textwidth]{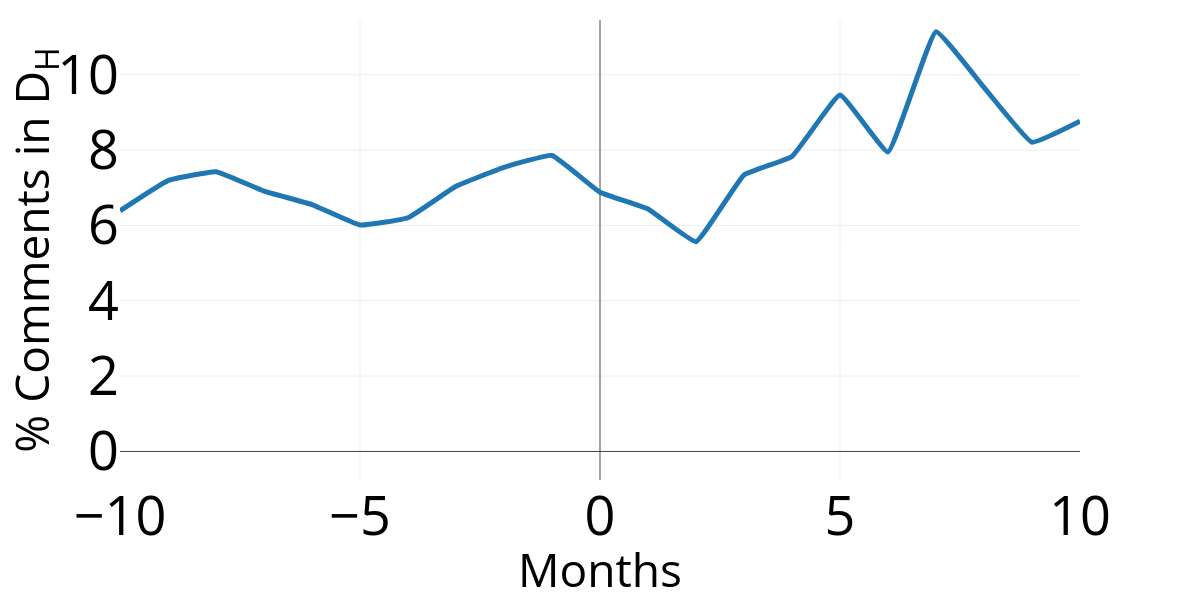}
        \caption{Metric: Rate of participation in hate communities
        ($\mathfrak{D}_H$). Event: Banning dangerous subreddit ($\mathfrak{D}_B$).}
        \label{fig:moderation:results:banning:community}
    \end{subfigure}
    \begin{subfigure}{.235\textwidth}
        \includegraphics[width=\textwidth]{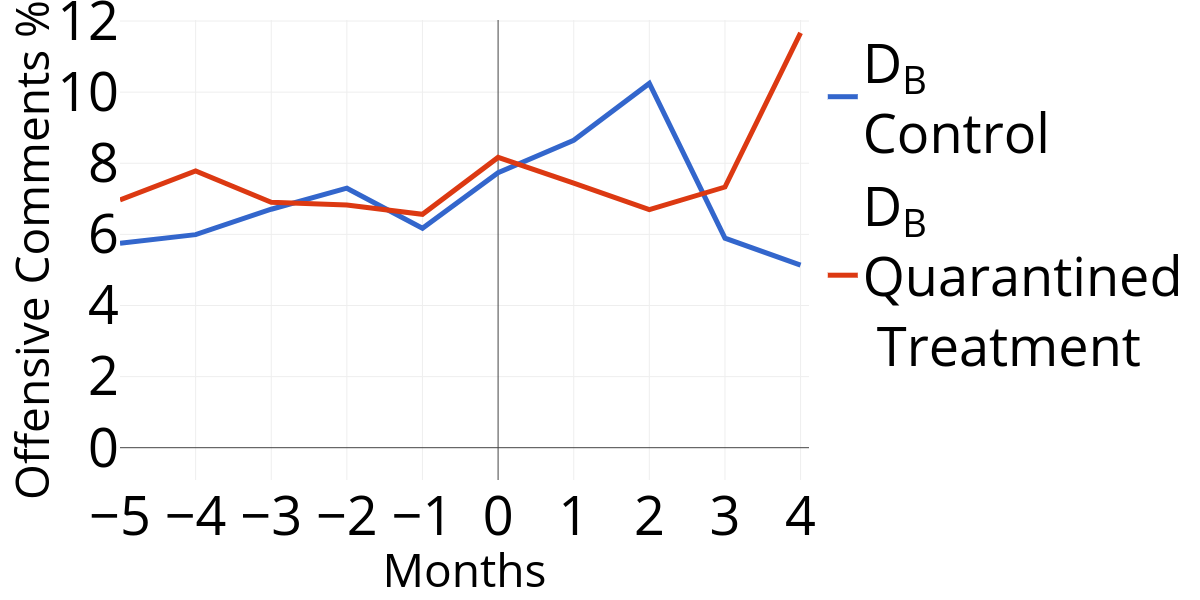}
        \caption{Metric: Offensiveness rate. Event: Quarantining dangerous subreddit
        ($\mathfrak{D}_B$).}
        \label{fig:moderation:results:quarantining:offensiveness}
    \end{subfigure}
    \begin{subfigure}{.235\textwidth}
        \includegraphics[width=\textwidth]{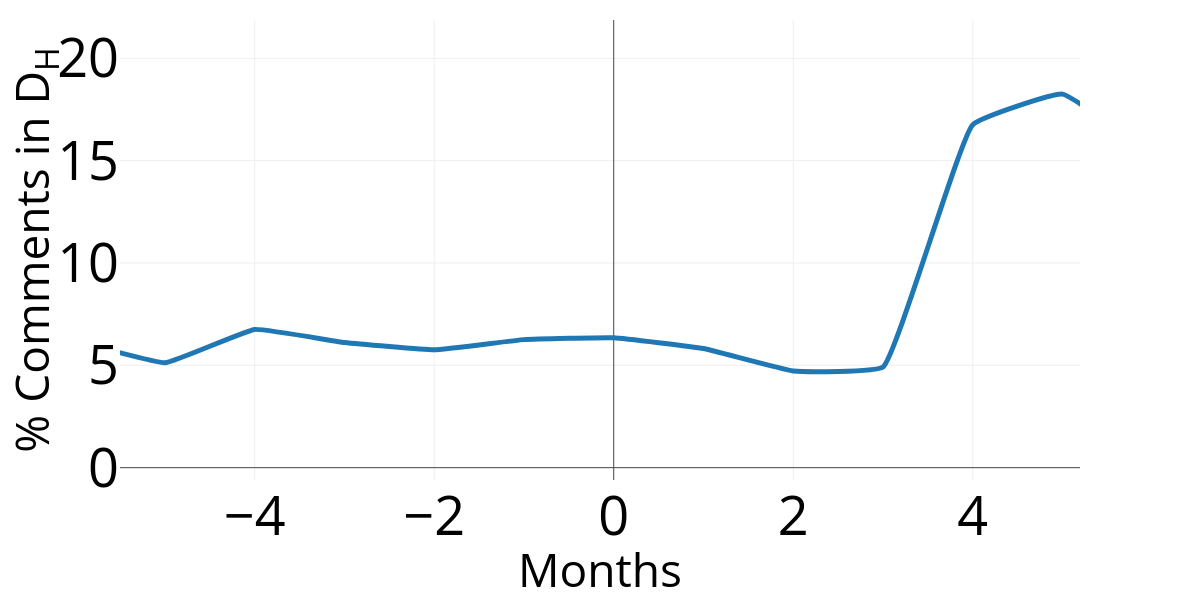}
        \caption{Metric: Rate of participation in hateful communities
        ($\mathfrak{D}_H$). Event: Quarantining dangerous subreddit ($\mathfrak{D}_B$).}
        \label{fig:moderation:results:quarantining:community}
    \end{subfigure}
    \caption{The impact of ban and quarantine events on user behavior. Month
    ``0'' represents the month of the event.}
  \label{fig:moderation:results:intervention}
\end{figure*}

\Cref{fig:moderation:results:intervention} shows the impact of banning or
quarantining communities on the behavior of their members. We make the
following observations.

\para{Bans and quarantines of dangerous communities do not reduce
offensiveness of the impacted members.} Looking at
\Cref{fig:moderation:results:banning:offensiveness} and
\Cref{fig:moderation:results:quarantining:offensiveness}  we can see that there
is no significant difference in incidence rates of offensive comments before
and after a ban or quarantine event for our treatment users. 

\para{Impacted community members of bans and quarantines simply move to other
hate subreddits.} We see from \Cref{fig:moderation:results:banning:community}
that users impacted by community bans continue a high level of engagement in
other hate communities. Surprisingly, our data shows
(\Cref{fig:moderation:results:quarantining:community}) that when communities
get quarantined, their users typically increase their participation in hate
communities significantly -- after a lag of several months (and possibly after
the quarantined community was banned).

\subsubsection{Takeaways: Do ban and quarantine interventions result in
improved user behavior?} \label{sec:moderation:results:takeaways}

Taken together, our analysis shows that participation in dangerous
($\mathfrak{D}_B$) and hateful ($\mathfrak{D}_H$) subreddits does have an
impact on a users behavior in the wider community and current administration
strategies of bans and quarantines are not effective for mitigating the impact
of this participation on individual users. Through our results, we are able to:
(1) \emph{confirm hypothesis (H3a) -- \ie user participation in hateful
subreddits negatively changes the nature of their participation in the broader
community} and (2) \emph{reject hypothesis (H3b) -- \ie we see that banning or
quarantining subreddits does not result in improved user behavior.} This
suggests the need for more nuanced interventions to mitigate the impact of
hateful and dangerous communities.

\section{Related Work}\label{sec:related}

At a high-level, in this work, we make contributions in three dimensions:
First, we perform measurements to understand how topics and user bases of
online communities change over time (\Cref{sec:evolution}).
Second, we identify the predictors of ``\emph{negative evolution}'' of
a community -- \ie evolution into an uncivil or offensive community
(\Cref{sec:predictors}). 
Finally, we conduct large scale measurements to understand how user behavior
can be influenced by moderator actions which enable or disable participation
within toxic communities (\Cref{sec:moderation}). 
In this section, we break down the related work in each of these dimensions.
Specifically, in \Cref{sec:related:evolution} we explore related research
seeking to understand how communities evolve and in
\Cref{sec:related:incivility} we explore research aimed at measuring and
mitigating incivility in online communities.

\subsection{Evolution of online communities}\label{sec:related:evolution}

Participation in online communities is increasingly common and studying
behavioral patterns and evolution in these communities has been the subject of
several research efforts. These efforts can be taxonomized by whether the goal
is to understand evolution of interaction quantity or quality. Below, we
present (a subset) of related work in these categories. 

\para{Interaction quantity.} Research in this category has generally focused on
understanding how the amount of interaction occurring in a community changes
over time and under different conditions. A general approach is to model
community interactions as a network graph where edges denote interactions (\eg
messages sent between two users) between nodes (\ie community members) and
track their evolution under different conditions. Especially relevant to our
work is research from Crandall \etal \cite{Crandall-KDD2008} which among other
results showed that \emph{interaction network related features are predictive
of future user behavior in topic-centered communities}. 
Other research focused
exclusively on understanding (rather than predicting) user interaction
evolution in different communities. Kumar \etal \cite{Kumar-LM2010} studied the
Flickr and Yahoo! 360 communities to understand the role of specific users in
community growth. They found that a small number of key users are responsible
for expanding a community and in the absence of these groups community growth
stagnated. Ngamkajornwiwat \etal \cite{Ngamkajornwiwat-HICSS2008} focus on
understanding how the network structure of online open source software
development communities evolve over time. They found that relationships
between ``\emph{core}'' and ``\emph{peripheral}'' developers generally weakened
over time and contributions from the peripheral members tended to decrease.
\emph{These results generally highlight the influence of a few community
members on the direction and trends observed in a community}. 

\para{Interaction quality.} Research in this category has generally focused on
understanding how the nature of interactions (\ie its qualities) within
a community change over time and under different conditions. 
Garcia \etal \cite{Garcia-COSN2013}, in a post-mortem of Friendster, showed
that {it is insufficient to consider only interaction network related
features when modeling a community's \emph{resilience} to decline specifically
highlighting the need to consider qualitative and external features}. The
importance of external events is also highlighted by Zannettou \etal
\cite{Zannettou-IMC2018, Zannettou-IMC2017} who focused on the evolution of
memes and news sources within communities and uncovered their influence on
external communities. Focusing exclusively on Reddit, Mills \etal
\cite{Mills-AIS2018, Mills-SCSM2015} showed that, for \subreddit{The\_Donald}
and \subreddit{Sanders4President}, external events and their community
participation guidelines were largely responsible for their rise in popularity
and large influx of users. \emph{These studies highlight the need to consider
cross-community interactions and external events when considering evolution of
communities}.
Several studies have also investigated how specific user interactions are
influenced by the age of a community. Danescu \etal \cite{Danescu-WWW2013}
found that linguistic features in a community were constantly evolving and
found that its newest members were most likely to adapt their own linguistic
features to those of the community. Gazan \cite{Gazan-HICSS2009} found that,
when communities stabilized, topics tended to move away from topical and
factual to personal and social. This generally resulted in increased
participation, often at the cost of conflict and factionalism. Rather
surprisingly, Kiene \etal \cite{Kiene-CHI2016} showed that after a certain
point in the life-cycle of a community, large influxes of users had no impact
on the quality of discourse within the subreddit. \emph{These studies highlight
the need to consider age and stability of a community when predicting its
evolution}.
Focusing on the impact of \emph{status} in online communities,
Bhalla \etal \cite{Bhalla-CMC2007} found that, over time, qualitative
characteristics of a community began matching the characteristics of \emph{the
elite}. Along similar lines, Gervais \cite{Gervais-PC2014} and Kwon \etal
\cite{Kwon-IR2017} showed that exposure to incivility from political elites
results in more  offensive rhetoric in online communities. In terms of methods,
we find most similarity between our approach and the work of Matias
\cite{Matias-CHI2016, Matias-2016} which used a logistic regression model to
attribute weights to survey-derived features to uncover the factors associated
with moderators and subreddits participating in the Reddit-wide blackout of
2015 -- in protest of Reddit's administrative actions. They uncovered a strong
correlation between moderator participation in meta-reddit subreddits and
community participation in the protest. \emph{These findings further highlight
the important role played by a few key members (elites and moderators) in
a community}.  

 
\subsection{Incivility in online communities}\label{sec:related:incivility}

Online communities have become an increasingly popular medium for different
types of discourse. This is particularly true for discourse around
controversial sociopolitical topics due to the pseudonomity provided by the
Internet and the resulting disinhibition effect \cite{Suler-CB2004}. Analyzing 
this discourse has been the subject of much research, with many focusing on
``incivility'' which is defined as ``communication that violates the norms of
offensiveness'' \cite{Mutz-2015}. These research efforts can be taxonomized
by whether the goal is to measure the incidence rates and propagation of
incivility or to identify and evaluate strategies for mitigating the impact of
such behaviors. Below, we highlight (a subset) of related work in these
categories.

\para{Measuring incivility.} Research in this category has generally focused on
understanding the incidence rates and propagation of online offensive discourse
in different contexts. Olteanu \etal \cite{Olteanu-ICWSM2018} found that
offensive speech on Reddit and Twitter were correlated with the occurrence of
violent offline events. Along similar lines, Nithyanand \etal
\cite{Nithyanand-Arxiv2017, Nithyanand-FOCI2017} found that external political
events and a change in media consumption habits in online communities were
correlated with a rapid increase of online incivility in political subreddits.
In addition, they traced the movement of offensive authors across communities
highlighting the sources and sinks of offensiveness on Reddit.

\para{Mitigating incivility.} Research in this category has generally focused
on understanding the impact of different types of interventions on uncivil
behaviour. Massanari \cite{Massanari-NMS2017} conducted a qualitative analysis
of the Reddit communities at the center of the Fappening
\cite{Fappening-Media2014-1, Fappening-Reddit2014-1} and Gamergate  
controversies \cite{Gamergate-Media2014-1, Gamergate-Media2014-2}. The study
highlights how the inaction of Reddit administrators and community moderators
resulted in the emergence of toxic technocultures and argues for the
exploration of alternative designs and moderation tools to combat the spread of
such toxicity. This study is complemented by Lapidot \etal
\cite{Lapidot-CHB2012} which demonstrated that, given anonymity, there was
a strong positive correlation between supervision and toxic inhibition.
\emph{These studies demonstrate that the absence of interventions (\ie
moderation) is correlated with incivility.} 
Birman \cite{Birman-GaTech2018} and Fiesler \etal \cite{Fiesler-ICWSM2018}
conduct measurements and characterize the different (explicitly stated)
moderation strategies employed by different communities on Reddit.
Complementing these works, Eshwar \etal \cite{Eshwar2018} study
moderator-removed comments and use this information to understand the
(implicit) moderation strategies and subreddit rules. Along similar lines,
Kiene \etal \cite{Kiene-CHI2016} analyzed moderator behavior during large
influxes of users into a community to uncover strategies for maintaining
content quality and civility. 
In terms of methods, we find most similarity between our work and research from
Chandrasekharan \etal \cite{Chandrasekharan-CSCW2017}. They authors focus on
uncovering the effectiveness of two subreddit bans: \subreddit{FatPeopleHate}
and \subreddit{CoonTown} by tracking the behaviour of two sets of users --
a treatment set (who visited the banned subreddits) and a control set (who
shared identical subreddit memberships as the treatment, except for presence in
the banned subreddits) before and after the bans. The study concluded that the
bans resulted in a general reduction of offensiveness in all interactions for
members belonging in the treatment group. 
In our work, we follow an identical approach only differing in the scale and
points of interest. First, our study considers over 3K subreddits including
100's of offensive, banned, and quarantined subreddits. Second, we are
interested in analyzing how user behaviour changes as a consequence of two
events: (1) when they join a community and (2) when the community is banned or
quarantined.
\emph{Taken together, these studies enumerate the different moderation
strategies used by online communities and methods for assessing their
effectiveness}.

\section{Conclusions}\label{sec:conclusions}

In this paper, we conducted a comprehensive study on the feasibility of
proactive moderation strategies for Reddit communities. Specifically, we showed
that Reddit communities, on average, have a high rate of evolution and this
requires constant monitoring, from moderators and administrators, for hateful
and dangerous communities -- a prohibitively expensive proposition
(\Cref{sec:evolution}). However, our observation that the evolutionary patterns
for hateful communities (\ie those frequently reported for hate speech) and
dangerous communities (\ie those which were eventually banned by Reddit for
violating the content policy) are different from other communities yields an
opportunity for easing moderator effort. We harness this insight to build
simple and explainable machine learning models which are able to study
evolutionary characteristics of subreddits and predict their future behavior
with reasonably high accuracy. Each of our classifiers show that structural
properties of the community (\ie which other communities its users interact
with) are the strongest predictors of future community behavior
\Cref{sec:predictors}. This finding suggests that tools which capture the
connectivity of subreddits to known hate or banned subreddits can be used to
identify which subreddits require careful monitoring or pre-emptive
interventions. Finally, we consider the impact of different events on user
behavior and show that joining hateful subreddits significantly worsens user
civility even in the broader context. Also, worryingly we find that current
methods for community-level interventions (\ie bans and quarantines) do not
have an impact on the civility of their impacted users (\cref{sec:moderation}).
This suggests that there is a need for the development of more active and
nuanced intervention strategies to effectively moderate hateful and dangerous
communities. While our results are largely focused on Reddit, one of the
largest and most controversial online communities, our findings can be useful
for administrators and moderators of any online community considering proactive
moderation to prevent the growth of hateful and dangerous communities.

\balance


\bibliographystyle{ACM-Reference-Format}
\bibliography{reddit}



\end{document}